\documentclass[aps,prb,twocolumn,groupedaddress,showpacs]{revtex4-1}

\usepackage{graphicx}
\usepackage{amsmath}
\usepackage{color}
\usepackage{hyperref}

\begin{document}

\title{The electronic specific heat of Ba$_{1-x}$K$_x$Fe$_2$As$_2$ from 2K to 380K.}

\author{J.G. Storey$^{1,2}$, J.W. Loram$^2$, J.R. Cooper$^2$, Z. Bukowski$^3$ and J. Karpinski$^3$}

\affiliation{$^1$School of Chemical and Physical Sciences, Victoria University, P.O. Box 600, Wellington, New Zealand}

\affiliation{$^2$Cavendish Laboratory, University of Cambridge, Cambridge CB3 0HE, U.K.}

\affiliation{$^3$Laboratory for Solid State Physics, ETH Zurich, Zurich, Switzerland}

\date{\today}

\begin{abstract}
Using a high-resolution differential technique we have determined the electronic specific heat coefficient $\gamma(T)=C^{el}/T$ of Ba$_{1-x}$K$_x$Fe$_2$As$_2$ with $x$ = 0 to 1.0, at temperatures ($T$) from 2K to 380K and in magnetic fields $H$ = 0 to 13T. In the normal state $\gamma_n(x,T)$ increases strongly with $x$ at low temperature, compatible with a mass renormalisation  $\sim12$ at $x$ = 1, and decreases weakly with $x$ at high temperature. A superconducting transition is seen in all samples from $x$ = 0.2 to 1, with transition temperatures and condensation energies peaking sharply at $x$ = 0.4. Superconducting coherence lengths $\xi_{ab}\sim20$\AA\ and $\xi_c\sim3$\AA\ are estimated from an analysis of Gaussian fluctuations. For many dopings we see features in the $H$ and $T$-dependences of $\gamma_s(T,H)$ in the superconducting state that suggest superconducting gaps in three distinct bands. A broad ``knee'' and a sharp mean-field-like peak are typical of two coupled gaps. However, several samples show a shoulder above the sharp peak with an abrupt onset at $T_{c,s}$ and a $T$-dependence $\gamma_s(T)\propto \sqrt{1-T/T_{c,s}}$. We provide strong evidence that the shoulder is not due to doping inhomogeneity and suggest it is a distinct gap with an unconventional $T$-dependence $\Delta_s(T)\propto (1-T/T_{c,s})^{0.75}$ near $T_{c,s}$. We estimate band fractions and $T$ = 0 gaps from 3-band $\alpha$-model fits to our data and compare the $x$-dependences of the band fractions with spectroscopic studies of the Fermi surface. 
\end{abstract}


\maketitle

The electronic specific heat $\gamma$ measured to room temperature contains a wealth of quantitative information about the electronic spectrum of metallic systems over an energy region $\pm$100meV about the Fermi level, crucial for understanding high-temperature superconductivity.
Measurements of the electronic specific heat have played an important role in revealing key properties of the copper-oxide based `cuprate' high-temperature superconductors (HTSCs). Some examples include the normal-state ``pseudogap''\cite{OURWORK2,LORAMPG,LORAMU0,ENTROPYDATA2}, the bulk sample inhomogeneity length scale\cite{SCFLUC}, and more recently evidence that the superconducting transition temperature is suppressed due to superconducting fluctuations\cite{TALLONFLUCS}. 
In this work we extend such measurements to the iron-arsenide based `pnictide' HTSCs. Here we present results obtained for polycrystalline samples of Ba$_{1-x}$K$_x$Fe$_2$As$_2$ ($x$ = 0 to 1.0) using a high-resolution differential technique\cite{LORAMCALORIMETER}.

With this technique we directly measure the difference in the specific heat capacities of a doped sample and an undoped reference sample (BaFe$_2$As$_2$). This eliminates most of the large phonon term from the raw data and yields a curve dominated by the difference in electronic terms.
Features of the electronic specific heat that would otherwise be masked by the large phonon background are then clearly visible in the raw data over the entire temperature range. Central to the success of this technique are measurements on a series of samples at closely spaced doping intervals, so that systematic trends in the the relatively small difference in phonon terms between sample and reference can be identified and appropriate corrections made. After making these corrections, the differences in electronic specific heat are determined with a resolution of $\sim$ 0.1 mJ mol$^{-1}$ K$^{-2}$, i.e. up to 1 in 10$^4$ of the total heat capacity, at temperatures from 2K to 380K and in magnetic fields from 0T to 13T. During a measurement run the total specific heats of the sample and reference are also measured. Using information on the electronic contribution deduced from our differential measurements we are able to extract the total phonon contribution from the total specific heat with a high degree of confidence, and crucially without recourse to arbitrary fitting and extrapolation procedures - a severe shortcoming of conventional heat capacity measurements in this temperature range. This allows us to determine a harmonic phonon spectrum which we compare with inelastic neutron data.

\section{Sample Preparation.}
\begin{figure}
\centering
\includegraphics[width=\linewidth]{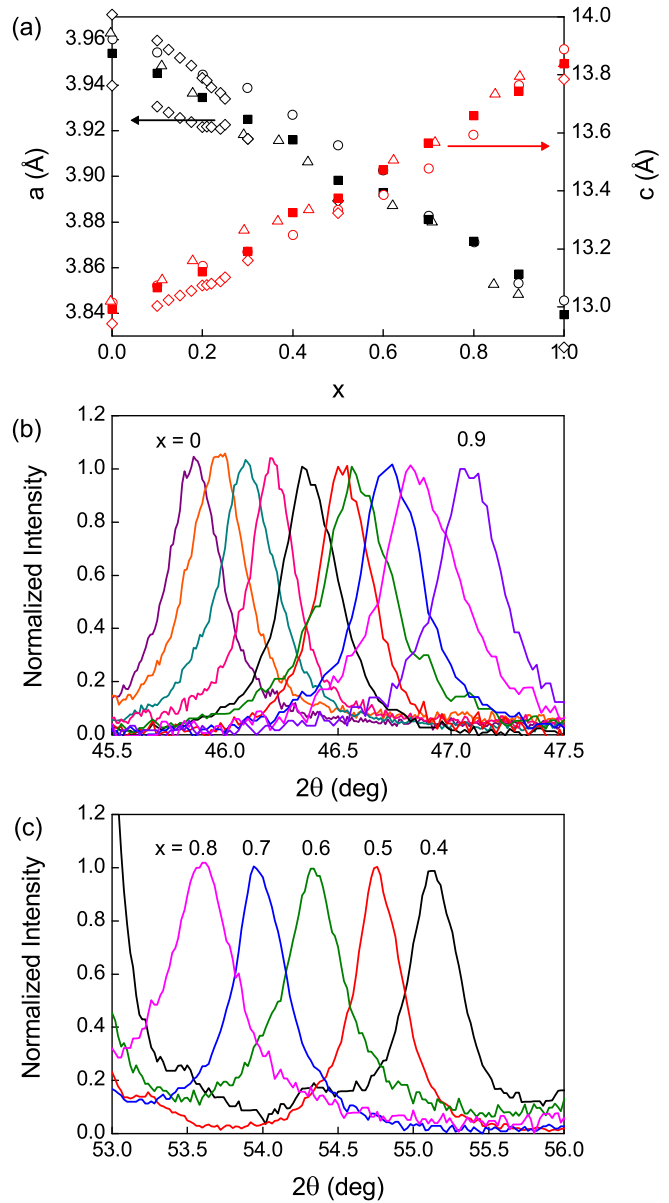}%
\caption{
(Color online) (a) The lattice parameters at room temperature as a function of nominal composition (solid squares). The open data points for comparison are from Chen \textit{et al}.\cite{CHEN} (circles), Johrendt \& P\"ottgen\cite{JOHRENDT} (triangles), and Avci \textit{et al.}\cite{AVCI2} (diamonds) whose data includes $b$-axis parameters in the orthorhombic phase. X-ray diffraction peaks corresponding to the (b) (2,0,0) and (c) (1,0,7) reflections. The peaks are symmetric and there is little variation in the widths with $x$.
} 
\label{LATTICEFIG}
\end{figure}

Polycrystalline samples of Ba$_{1-x}$K$_x$Fe$_2$As$_2$ were prepared by a solid state reaction method similar to that reported by Chen \textit{et al}\cite{CHEN}. First, Fe$_2$As, BaAs, and KAs were prepared from high purity As (99.999\%), Fe (99.9\%), Ba (99.9\%) and K (99.95\%) in evacuated quartz ampoules at 800, 650 and 500$^\circ$C respectively. Next, the terminal compounds BaFe$_2$As$_2$ and KFe$_2$As$_2$ were synthesized at 950 and 700$^\circ$C respectively, from stoichiometric amounts of BaAs or KAs and Fe$_2$As in alumina crucibles sealed in evacuated quartz ampoules. Finally, 11 samples of Ba$_{1-x}$K$_{x}$Fe$_2$As$_2$ with $x$ = 0 to 1.0 were prepared from appropriate amounts of single-phase BaFe$_2$As$_2$ and KFe$_2$As$_2$. The components were mixed, pressed into pellets, placed into alumina crucibles and sealed in evacuated quartz tubes. The samples were annealed for 50 h at 700$^\circ$C with one intermediate grinding, and were characterized by room temperature powder X-ray diffraction using Cu K$_\alpha$ radiation. The diffraction patterns were indexed on the basis of the tetragonal ThCr$_2$Si$_2$ type structure (space group I4/mmm). Lattice parameters calculated by a least-squares method agree well with those reported by Chen \textit{et al}.\cite{CHEN}, Johrendt and P\"ottgen\cite{JOHRENDT} and Avci \textit{et al.}\cite{AVCI2} as illustrated in Fig.~\ref{LATTICEFIG}. The linear changes in lattice parameters with $x$ suggest that there is little difference between the nominal and actual K concentration in the samples. Note that Avci \textit{et al.}\cite{AVCI2} estimate a compositional uncertainty $\Delta x\lesssim 0.01$ from inductively coupled plasma elemental analysis. The samples for heat capacity measurement were cut from the larger pellets using a diamond wheel saw, and weighed approximately 0.8g. All samples were stored in an argon atmosphere, cut under flowing argon and exposed only briefly to air, for less than 30 minutes  while mounting  them in the calorimeter or SQUID magnetometer.

\section{Phonon Correction.}
\begin{figure}
\centering
\includegraphics[width=\linewidth]{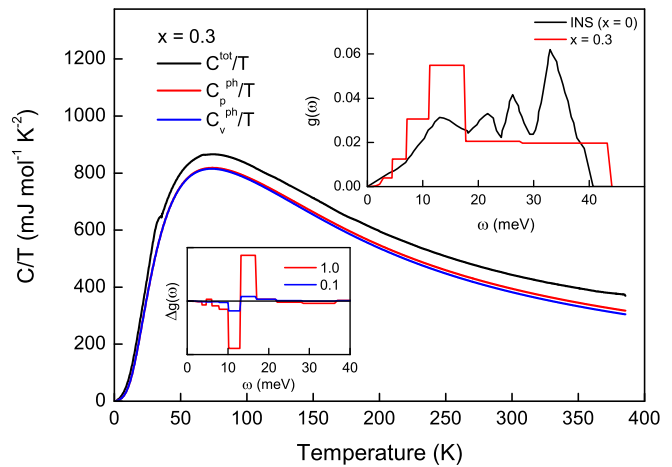}%
\caption{
(Color online) Total specific heat of Ba$_{0.7}$K$_{0.3}$Fe$_2$As$_2$ and phonon terms at constant pressure $C_p^{ph}/T=C^{tot}/T-\gamma$ and constant volume $C_v^{ph}/T=C_p^{ph}/T-\gamma_{an}$. Upper inset: Phonon density of states histogram extracted from $C_v^{ph}/T$ for $x$ = 0.3, compared with the phonon spectrum of BaFe$_2$As$_2$ measured by inelastic neutron scattering\cite{MITTAL}. The curves are normalised by area. Lower inset: Difference in phonon density of states between $x$ = 0.1 and 0, and between 1.0 and 0.
} 
\label{GPHFIG}
\end{figure}

\begin{figure}
\centering
\includegraphics[width=\linewidth]{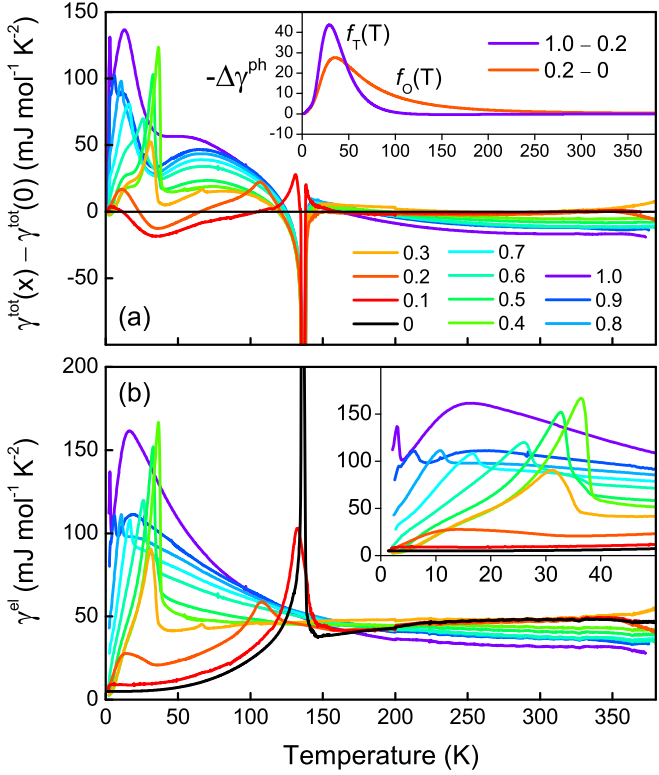}%
\caption{
(Color online) (a) The directly measured raw difference in specific heat coefficients between the doped samples and undoped reference. The inset shows the temperature dependence of the residual phonon term corrections for the tetragonal (T) and orthorhombic (O) phases. (b) The electronic specific heat of Ba$_{1-x}$K$_x$Fe$_2$As$_2$ up to 380K (main figure) and 50K (inset). 
} 
\label{GAMMAFIG}
\end{figure}

The total specific heat coefficient $\gamma^{tot}=C^{tot}/T$ can be written as $\gamma^{tot}=\gamma+\gamma^{ph}+\gamma^{an}$, where $\gamma$, $\gamma^{ph}$ and $\gamma^{an}$, represent the electronic, phonon and anharmonic contributions respectively (see Fig.~\ref{GPHFIG}). Here $\gamma^{ph}=C_v^{ph}/T$ and $\gamma_{an}=C_p^{ph}/T-C_v^{ph}/T$.
Differential measurements of Ba$_{1-x}$K$_x$Fe$_2$As$_2$ (see Fig.~\ref{GAMMAFIG}(a)) between each sample and the $x$ = 0 reference give $\Delta\gamma^{tot}=\Delta\gamma+\Delta\gamma^{ph}$ (assuming that $\Delta\gamma^{an}$ = 0), where $\Delta\gamma^{ph}$ is a bell shaped curve peaking typically at around 30 - 40K and varying as $ T^2$ at low temperatures and $ 1/T^3$ at high temperatures\cite{LORAM93}. Within a single crystallographic phase $\Delta\gamma^{ph}(x,T)$ is generally found to be a separable function of $x$ and $T$, i.e. $\Delta\gamma^{ph}(x_2-x_1,T)=(x_2-x_1)f(T)$, and is thus a universal function of $T$, scaling in magnitude with changes in $x$ \cite{LORAM93}. A simple calculation shows that this separable form is expected for phonon shifts of up to $\sim10\%$. If the shifts have a linear doping dependence $\Delta\omega/\omega = a(\omega)\Delta x$, then $f(T)= -\int{a(\omega)(\partial C_E/\partial T)_\omega g(\omega)d\omega}$, where $C_E(\omega/T)$ is the Einstein specific heat function for a single harmonic oscillator and $g(\omega)$ is the phonon density of states. Higher order corrections $\propto(\Delta x)^2$ will only arise if $\Delta \omega/\omega$ is non-linear in $\Delta x$, or for very large frequency shifts. This simple separable behaviour greatly increases the reliability of the correction for $x$-dependent changes in phonon specific heat.
 To correct $\Delta\gamma^{tot}$ for the doping dependent changes in the phonon term we must first determine $f(T)$. Inspection of Fig.~\ref{GAMMAFIG}(a) reveals a systematic negative peak at 35K which grows with $x$, and is consistent with an increase in phonon frequencies expected from the substitution of heavier Ba by lighter K. A simple estimate of average fractional phonon shifts using the approximate relation $\Delta\omega/\omega\sim-
 \Delta\gamma^{ph}/(dC^{ph}/dT)$ yields $\Delta\omega/\omega\sim 2.5\%$ over the range $0\leq x< 0.3$ and $\sim 4\%$ over the range $0.3\leq x\leq 1$. A suitable $f(T)$ was constructed to remove this peak for each phase, the criterion being that after applying the correction, no evidence for the broad 35K peak should be visible in any of the corrected curves. For the tetragonal phase $0.3\leq x\leq 1$ this  was achieved by scaling the correction curve $f_T(T)$, shown in the inset to Fig.~\ref{GAMMAFIG}(a), linearly with doping. In the magnetically ordered orthorhombic phase $0\leq x < 0.3$, the phonon correction $f_O(T)$ has a $T$-dependence similar to $f_T(T)$, but scales sub-linearly with increasing $x$. This sub-linear doping dependence correlates with the decrease of the spin-density-wave (SDW) order parameter with $x$, suggesting that magneto-phonon coupling is important in the magnetically ordered phase. To ensure that each resulting $f(T)$ has a $T$-dependence compatible with that of a phonon spectrum it was modelled with a histogram for the difference phonon spectrum with a fixed fractional bin width $\Delta\omega/\omega=0.238$, as discussed previously\cite{LORAM93}.

After applying the phonon correction $\Delta\gamma^{ph}(x,T)$ to $\Delta\gamma^{tot}(x,T)$ for all our samples we obtain the difference of electronic terms $\Delta\gamma(x,T)= \gamma(x,T)-\gamma(0,T)$ between $x$ and $x$ = 0. To obtain the electronic term for each sample from this differential data requires that $\gamma(T)$ for one sample is known or assumed, and for this we choose the $x$ = 0.3 sample. After removing the broad negative peak in $\Delta\gamma^{tot}$ at 35K for this sample $\Delta\gamma(0.3,T)$ has an additional negative $T$-dependence given by $\sim$ $-20\times10^{-6}T^3$  mJ/mol K$^2$ in the range 40 to 110K. The negative curvature of this term, already evident above 80K in the raw $\Delta\gamma^{tot}$ data shown in Fig.~\ref{GAMMAFIG}(a), continues to increase in magnitude up to 136K and then abruptly vanishes at the magneto-structural transition. For this reason we associate it with the $T$-dependence of the electronic and magnetic (magnon) specific heat coefficient $\gamma(x=0)$ of the undoped sample in the SDW phase.
From $\Delta\gamma(0.3,T)$ a roughly $T$-independent normal-state $\gamma_n(0.3)$ below $T_c$ of 47mJ mol$^{-1}$ K$^{-2}$ is inferred from the entropy conservation constraint between normal and superconducting states.
%, together with a relatively weak $\sim T^3$ increase in $\gamma(0,T)$ below 35K for $x$ = 0. 
Finally, we note that the low temperature value for $\gamma_n(0.3)$ is very close to its high temperature value $\sim$ 50mJ mol$^{-1}$ K$^{-2}$ determined directly from the difference between $\gamma^{tot}(0.3)$ and the saturation value of $\gamma^{ph}=3nR$, where $n=5$ is the number of atoms per formula unit. We therefore assume that $\gamma_n(0.3)$ is approximately $T$-independent over the entire range, and choose the $\gamma(0.3,T)$ for $x$ = 0.3 to be the ``known reference''. The electronic term $\gamma(x,T)$ for all other samples is then calculated from $\gamma(x,T)=\Delta\gamma(x,T)-\Delta\gamma(0.3,T)+\gamma(0.3,T)$. Final curves for $\gamma(x,T)$ are shown in Fig.~\ref{GAMMAFIG}(b). It is important to note that any error in our assumption that $\gamma_n$ for $x$ = 0.3 is approximately $T$-independent will affect equally the resulting curves for $\gamma(T)$ for all other dopings and will have no effect on differences in $\gamma(T)$ between samples. The fact that all the curves for $\gamma(T)$ vary smoothly with temperature with no sign of the phonon correction or the large 135K anomaly present in the raw data (Fig.~\ref{GAMMAFIG}(a)) provides confidence in the corrections and procedure discussed above and in the accuracy and reproducibility of the raw differential data. 

With the electronic terms in hand we are able to determine the phonon terms in $\gamma^{tot}$ directly and without resorting to arbitrary polynomial fits. The total specific heat ($C^{tot}/T=\gamma^{tot}$) and phonon terms at constant pressure ($C_p^{ph}/T=\gamma^{tot}-\gamma$) and constant volume ($C_v^{ph}/T=C_p^{ph}/T-\gamma^{an}$) are shown in Fig.~\ref{GPHFIG} for $x$ = 0.3. 
The anharmonic term is given by\cite{ASHCROFT} $\gamma^{an}=VB\beta^2$ where $V$ is the molar volume, $B$ is the bulk modulus and $\beta$ is the volume expansion coefficient. We assume that $\gamma^{an}$ is doping independent and use $V$ = 61 cm$^3$ mol$^{-1}$ (Ref.~\onlinecite{ROTTER}), $B \sim$ 0.80$\times$10$^8$ mJ cm$^{-3}$ (Ref.~\onlinecite{KIMBER}), and $\beta$(300K) $\sim$ 50$\times$10$^{-6}$ K$^{-1}$ (Ref.~\onlinecite{BUDKO}) to obtain a room temperature value $\gamma^{an}$(300K) $\sim$ 12 mJ mol$^{-1}$ K$^{-2}$ for each sample.  To a very good approximation\cite{ASHCROFT} $\beta(T)\propto C_v^{ph}(T)$ and thus
$\gamma^{an}(T) = [C_v^{ph}(T)/C_v^{ph}(300\mathrm{K})]^2\times\gamma^{an}(300\mathrm{K})$. The upper inset in Fig.~\ref{GPHFIG} compares the phonon density of states extracted from $C_v^{ph}(x=0.3)/T$ with the spectrum for BaFe$_2$As$_2$ obtained from inelastic neutron scattering (INS) measurements by Mittal \textit{et al}.\cite{MITTAL} (We are unaware of INS data for any other K-doped samples). Our histogram reproduces the band width and the initial slope of the spectrum, though the neutron data places more weight at high frequencies. This is not unexpected as inelastic  neutron scattering measures a (generalised) phonon spectrum weighted by the scattering cross-sections and inverse masses of the different ions\cite{MITTAL,ZBIRI}, whilst the phonon specific heat gives equal weight to all phonon modes. The heavy Ba ions contribute almost entirely to the low frequency phonon spectrum below 20meV, but carry a low weighting in the neutron spectrum. This may account for the relatively greater weight in the phonon spectrum below 20meV and smaller weight at higher frequencies revealed by our measurements. In view of this, the agreement is quite reasonable and helps validate our main assumption that $\gamma_n(0.3,T)$ is approximately $T$-independent. 

\section{Electronic Specific Heat.}
\begin{figure*}
\centering
\includegraphics[width=\linewidth]{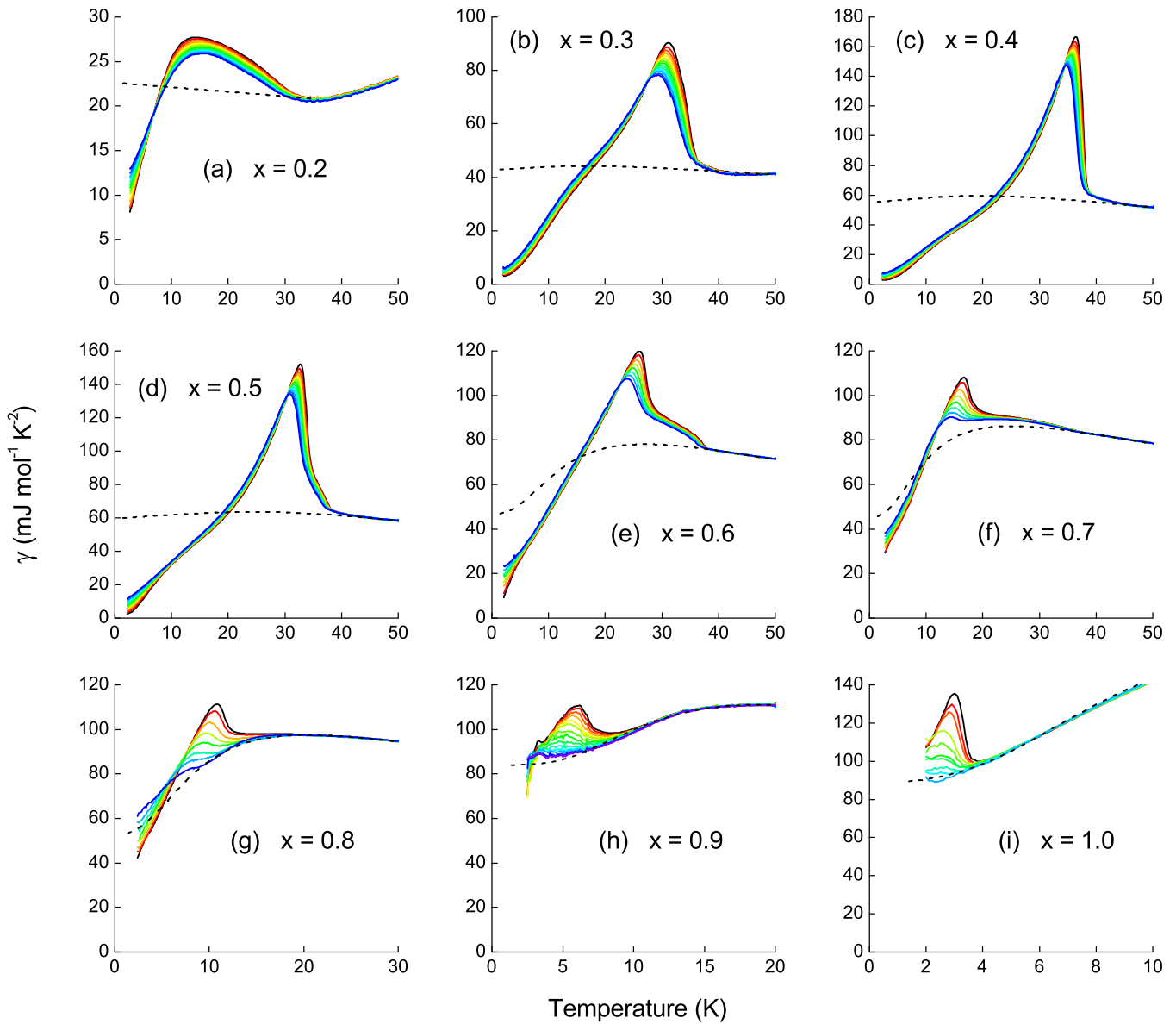}%
\caption{
(Color online) Electronic specific heat at zero (black line) and applied magnetic fields (colored lines) vs temperature. Entropy conserving normal-state curves are shown by the dotted lines. The applied fields shown are: 1 to 13T in 1T increments for $x$ = 0.2 to 0.4; 1 to 13T in 2T increments for $x$ = 0.6 to 0.8; 0.2, 0.5, 1.0, 1.5 then 2 to 11T in 1T increments for $x$ = 0.9; and 0.1, 0.2, 0.5, 0.7, 1.0, 1.2, 1.5, 2 and 3T for $x$ = 1.0.
} 
\label{GHVSTFIG}
\end{figure*}

\begin{figure}
\centering
\includegraphics[width=\linewidth]{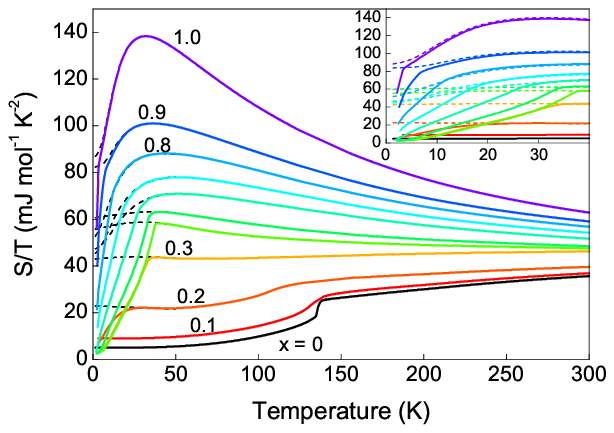}%
\caption{
(Color online) Temperature dependence of the electronic entropy for $x$ = 0 to 1.0 up to 300K (main figure)
and 40K (inset). Dotted lines show normal state curves.
} 
\label{ENTROPYFIG}
\end{figure}

Figure~\ref{GHVSTFIG} represents the first comprehensive high-resolution determination of the temperature, doping and magnetic field dependence of the absolute electronic specific heat of Ba$_{1-x}$K$_x$Fe$_2$As$_2$ across the entire series. This data together with Meissner effect measurements confirm that superconductivity is observed for $x\geq0.2$, and that the $x$ = 0 and 0.1 samples are non-superconducting at all temperatures. In Fig.~\ref{ENTROPYFIG} we show $S/T$ in zero field for all samples, where the electronic entropy $S(T) =\int_0^T{\gamma(T^\prime)dT^\prime}$. $S(T)/T$ is the average value of $\gamma(T)$ in the temperature range 0 to $T$, and equals $\gamma(0)$ at $T$ = 0.
Apart from $x$ = 0.9 and 1, the underlying normal state electronic term $\gamma_n$ below $T_c$ for the superconducting samples could not be determined directly by suppressing superconductivity with a magnetic field since their upper critical fields exceed 13T. However we can estimate the $T$-dependence of $\gamma_n$ below $T_c$ with reasonable confidence using the following constraints. (i) $\gamma_n$ and $S_n = \int{\gamma_n dT}$ are continuous with no change of slope through $T_c$, (ii) the normal state and superconducting entropies are equal at $T_c$, $S_n(T_c)=S_s(T_c)$ (i.e. the areas under $\gamma_n$ and $\gamma_s$ below $T_c$) are equal. For convenience we first choose a suitable $T$-dependence for $S_n/T$ and then obtain $\gamma_n$ from $\gamma_n=dS_n/dT$. The possible $T$-dependence for $S_n/T$ (and hence $\gamma_n$) is further restricted by the reasonable assumption for a Fermi liquid that $\gamma_n = \gamma_n(0)+bT^2$ (and $S_n/T=\gamma_n(0)+bT^2/2$) close to $T$ = 0. The broken lines in Figs.~\ref{GHVSTFIG} and \ref{ENTROPYFIG} show $\gamma_n$ and $S_n/T$ respectively for all the superconducting samples. As shown in Fig.~\ref{GHVSTFIG}  for $x\geq0.8$ we are able to significantly suppress the transition temperature in a field of 13T, and confirm our normal-state $\gamma_n$ curves to well below the zero field $T_c$. 

\begin{figure}
\centering
\includegraphics[width=\linewidth]{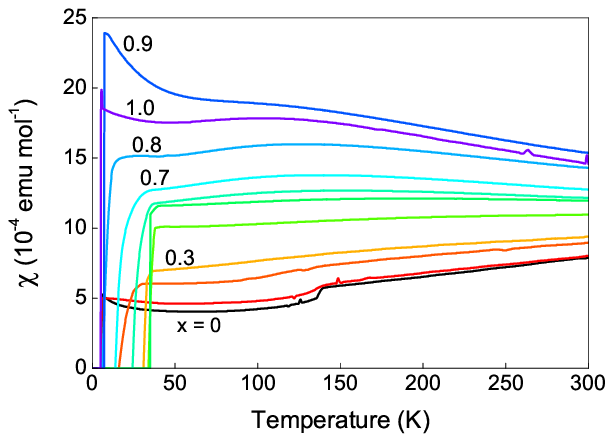}%
\caption{
(Color online) Temperature dependence of the static susceptibility measured at 3T for $x$ = 0 to 1.0.
} 
\label{CHIFIG}
\end{figure}

Figure~\ref{CHIFIG} shows the static susceptibility $\chi$ measured in 3T in a Quantum Design SQUID magnetometer.  $\chi(T)$ increases systematically with doping, apart from the $x$ = 0.9 data which lies above the $x$ = 1.0 data. This is probably due to the presence of a larger Curie-Weiss term in the $x$ = 0.9 sample. In their model, Kou \textit{et al}. predict a small  upturn at low-$T$ due to the presence of itinerant electrons coexisting with local, magnetically ordered moments\cite{KOU}. However the unsystematic variation in the size of this upturn, both in our data and in the literature\cite{ROTTER2,WANGXF}, suggests that it is predominantly extrinsic in nature. Above the SDW transition at $T_0$, $\chi$ increases linearly with $T$ for $x$ = 0 to 0.2, which has been attributed to antiferromagnetic correlations persisting above $T_0$\cite{ZHANG2,KOU}. For $x>0.4$, $\chi$ exhibits a broad peak between 100K and 200K and smoothly crosses over at higher temperatures to a decreasing Curie-like $T$-dependence. This is precisely the behaviour predicted by the quantum Heisenberg antiferromagnetic model\cite{ZHANG2,SU}. In a local moment model the steadily decreasing crossover temperature in our data implies that the ratio of the next-nearest to nearest neighbour magnetic super-exchange energies,  $J_2/J_1$, decreases with $x$.
On the other hand comparison of Figs.~\ref{ENTROPYFIG} and \ref{CHIFIG} shows that $S/T$ and $\chi$ have broadly similar $T$-dependences. This behaviour is typical of a Fermi liquid in which both properties are dominated by thermal excitation of quasi-particles.

\begin{figure}
\centering
\includegraphics[width=\linewidth]{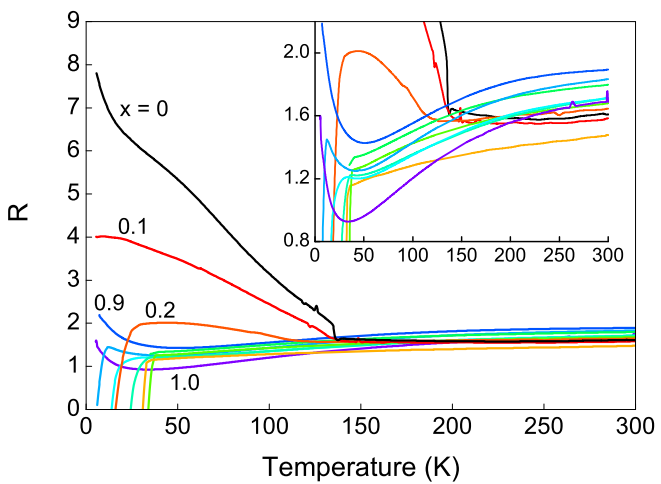}%
\caption{
(Color online) Temperature dependence of the Wilson ratio $R = R_0\chi T/S$, where $R_0=(\pi^2k_B^2)/(3\mu_B^2)$, for $x$ = 0 to 1.0.
} 
\label{WILSONFIG}
\end{figure}

The Wilson ratio $R = (\pi^2k_B^2\chi T)/(3\mu_B^2S)$ is plotted in Fig.~\ref{WILSONFIG}. Note that we have not corrected $\chi$ for $T$-independent core and Van Vleck terms which are unlikely to exceed $10^{-4}$ emu/mol. If the $g$-factor is 2, $R$ = 1 for non-interacting quasiparticles, 2 for spin-$\tfrac{1}{2}$ Kondo alloys and many heavy Fermion compounds and  using this definition of the Wilson ratio, 4.7 for non-interacting spin-$\tfrac{1}{2}$ moments. $R$ is expected to decrease with electron-phonon enhancement of $\gamma(T)$ and to increase with exchange enhancement of $\chi(T)$. For the entire doping range $0\leq x\leq 1$, $R$ increases weakly with increasing temperature in the paramagnetic tetragonal phase from $\sim$0.9 to 1.9 as shown in Fig.~\ref{WILSONFIG} inset. These values are reasonable for a correlated Fermi liquid. Just above $T_c$ the superconducting  samples with $x=0.2$ to 0.8 have $R$ = 1.2 to 1.3. In the orthorhombic SDW phase below $T_0$ for $x$ = 0 to 0.2, $R$ increases steeply with decreasing temperature since the entropy, determined solely by thermal excitations, falls more rapidly with increasing magnetic order than $\chi$. For $x=0$ the value $\chi(T=0)\simeq 4\times10^{-4}$ emu/mole in Fig.~\ref{CHIFIG} can be used to estimate the effective moment $p_{eff}$ per Fe atom in the SDW phase. Using the standard mean field formulae $\chi(T) = 2N_{av} p_{eff}^2/[3k_B(T+T_0)]$ and $\chi(0)\simeq \chi({T_0})$, where $N_{av}$ is Avogadro's number, we find $ p_{eff}=0.66\mu_B$ for $x$ = 0. The small step in $R$ at the magneto/structural transition $T_0$ can be explained if only $42\%$ of the entropy jump at $T_0$ is caused by the SDW transition and the rest arises from the structural transition which would not contribute to $\chi$. (This estimate uses the value $R$ = 1.6 found just above $T_0$ for $x$ = 0.)

\subsection{Normal State.}
We begin our discussion with the normal-state electronic specific heat $\gamma_n$, which exhibits a remarkable evolution with doping that is unlike anything we have seen previously in the cuprate high-temperature superconductors\cite{ENTROPYDATA2}.

In the $x$ = 0 sample there is a sharp and almost first order anomaly at the magneto-structural transition at $T_0$ = 136K, with a small second-order-like shoulder $\sim$1K above this. This value of $T_0$ agrees with published single crystal data\cite{DONG,ROTUNDU}. At low temperatures, $\gamma(0)\sim 4.6$ mJ mol$^{-1}$ K$^{-2}$ is a factor ten lower than its value above $T_0$, reflecting gapping, i.e. reconstruction, of the Fermi surface, and then increases as $\sim$ $20\times10^{-6}T^3$  mJ mol$^{-1}$ K$^{-2}$ in the range 40 $-$ 110K. This $T$-dependence is caused by quasiparticle and magnon excitations in the SDW phase. Band splitting and signs of partial gapping of the Fermi surface have been observed in the SDW state of BaFe$_2$As$_2$ by angle-resolved photoemission spectroscopy\cite{YANG}.
At low temperatures, $\gamma(0)=8.7$ mJ mol$^{-1}$ K$^{-2}$ for $x$ = 0.1, while for $x$ = 0.2, extrapolating from above the weak superconducting transition, $\gamma_n(0)$ = 22.3 mJ mol$^{-1}$ K$^{-2}$. This progressive increase in $\gamma(0)$ with $x$ arises from  the progressive reduction in the gap induced by SDW order and the larger number of free carriers at low $T$. For $x$ = 0.1 and 0.2 the corresponding anomalies at $T_0$ = 131.6K and 107K are broader and considerably reduced in magnitude compared with $x$ = 0, in agreement with the data of Rotter \textit{et al}.\cite{ROTTER} The magnetic field dependences of these anomalies are extremely weak. $^{57}$Fe-M\"{o}{\ss}bauer spectroscopy\cite{ROTTER} and neutron diffraction\cite{AVCI1,AVCI2} measurements show that the SDW phase is fully suppressed somewhere between $x$ = 0.2 and 0.3, and we see no evidence for a magneto/structural transition in our $x$ = 0.3 sample. The very weak anomaly at 67K in the differential data for this sample (Fig.~\ref{GAMMAFIG}(b)) is probably due to an FeAs impurity phase\cite{SELTE}. By comparing the height of this anomaly to that of pure FeAs\cite{GONZALEZ} we estimate a FeAs concentration of 4 mole \% in this sample. FeAs anomalies are absent in all of our other samples.
\begin{figure}
\centering
\includegraphics[width=\linewidth]{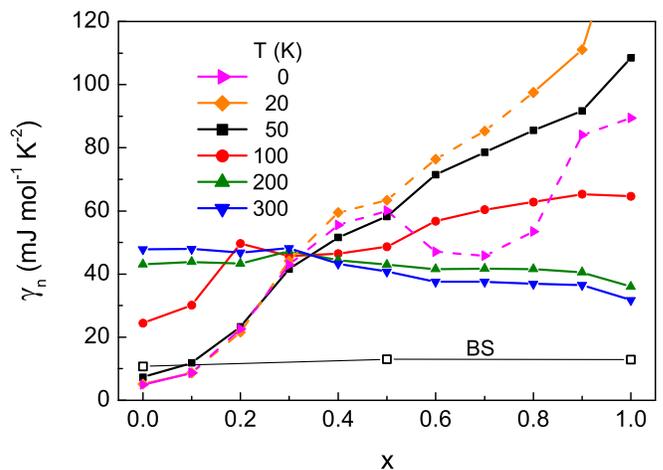}%
\caption{
(Color online) Normal state electronic specific heat $\gamma_n$ vs doping at several temperatures and band structure values\cite{SHEIN,HASHIMOTO2} for $x$ = 0, 0.5 and 1. Dotted lines show $\gamma_n$ for values of $x$ where $T<T_c$.
}
\label{GVSXFIG}
\end{figure}

Values of the normal state $\gamma_n$ at several fixed temperatures are shown in Fig.~\ref{GVSXFIG}. As $x$ increases from 0 to 0.3, the collapsing magneto-structural anomaly results in a gradual filling-in of the normal-state $\gamma_n$ at low-$T$, while above 200K $\gamma_n$ is only weakly dependent on $x$ and $T$ (see Figs.~\ref{GAMMAFIG}(b) \& \ref{GVSXFIG}). However for $x>0.3$ we observe a systematic decrease in $\gamma_n$ with $x$ and with $T$ at high-$T$. At low $T$  there is  a larger peak in $\gamma_n(T)$ that grows with $x$ and correlates with the growth in $\gamma_n(T=0)$, see Fig.~\ref{GAMMAFIG}(b). At intermediate concentrations $x$ = 0.5 to 0.8, this peak is masked by the superconducting transition, though its presence in the underlying normal state $\gamma_n$ can be inferred from the entropy conserving determinations of $\gamma_n$ and $S_n/T$ shown in Figs.~\ref{GHVSTFIG} and \ref{ENTROPYFIG}. The magnitude of the peak grows rapidly towards $x$ = 1, with the peak temperature falling with $x$ to 15K at $x$ = 1. A similar enhancement at low-$T$ and high-$x$ has also been observed in the related Sr$_{1-x}$K$_x$Fe$_2$As$_2$ system via conventional heat capacity techniques\cite{WEI}. Although we cannot entirely exclude the possibility that the low temperature peak in $\gamma_n$ is a result of an error in the phonon correction, we believe that this is unlikely for the following reasons. As noted above, the observed negative peak in the raw data at 35K, increasing progressively with $x$ across the entire series, is consistent with the expected increase in phonon frequencies on substituting heavy Ba with light K. To explain a positive peak, increasing non-linearly with $x$ at a temperature decreasing with $x$, in terms of phonons would require a substantial $T$ and $x$-dependent softening of low frequency phonon modes. We are unaware of any reason or evidence for such soft mode behaviour in heavily doped Ba$_{1-x}$K$_x$Fe$_2$As$_2$, and we believe this low-$T$ positive peak is more likely to be a feature of the electronic spectrum. 

\begin{figure}
\centering
\includegraphics[width=\linewidth]{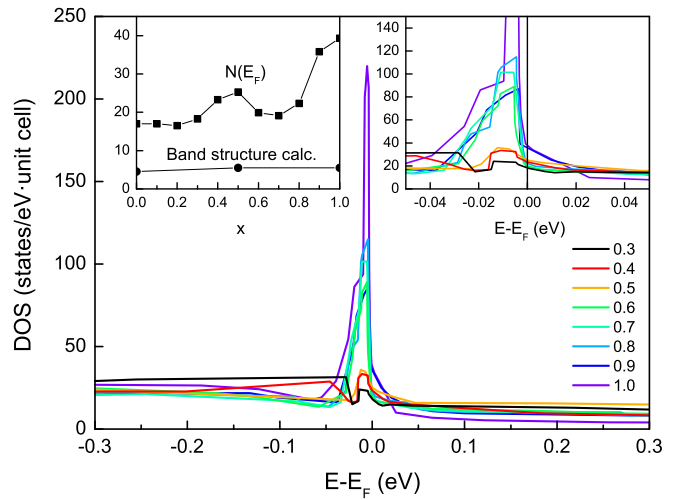}%
\caption{
(Color online) A set of model electronic density of states curves which reproduces $\gamma_n(x,T)$ from 0K to 380K.
The DOS at the Fermi level determined from $\gamma_n$ extrapolated to $T$ = 0 is compared with values from band structure calculations\cite{SHEIN,HASHIMOTO2} in the upper left inset.
} 
\label{DOSFIG}
\end{figure}

Also shown in Fig.~\ref{GVSXFIG} are band structure (bs) values\cite{SHEIN,HASHIMOTO2} for $\gamma_n$ for $x$ = 0, 0.5 and 1. Comparison with the experimental values show an electronic mass enhancement $m^*/m_{bs}$ of around 4 for $x$ = 0 ($T>T_0$) increasing to around 9 at 40K for $x$ = 1. This evolution is inconsistent with a simple shift of the Fermi level ($E_F$) towards a van-Hove singularity in the density of states (DOS) as proposed for example in overdoped cuprates\cite{STOREYENTROPY}.  This would give an increasing $\gamma_n$ with $x$ at all $T$ and a significant shift in the peak temperature of the low-$T$ hump in $\gamma_n(T)$ with $x$. 
Instead, the $x$-dependence at high and low-$T$ is suggestive of a transfer of spectral weight from high to low energies. This can been seen explicitly in a set of model effective DOS curves shown in Fig.~\ref{DOSFIG} that reproduces the $\gamma_n(x,T)$ data over the entire temperature range. In these calculations the chemical potential was adjusted in order to maintain the same number of quasiparticles at all $T$. The development of the low-$T$ hump between 15 and 20K requires a very sharp spike in the DOS located about 5meV from $E_F$, that grows in magnitude with $x$ at the expense of states on either side of it.  It is possible that this kind of structure arises from strong electron-electron correlations of the type that occur in heavy Fermion compounds.
The evolution of $\gamma_n(x,T)$ for $x>0.3$ can probably be explained in terms of a temperature dependent mass enhancement that increases with $x$, and indeed the $T$-dependence of $\gamma_n$ resembles the renormalization expected from coupling of the electrons to phonons\cite{GRIMVALL}, or to  paramagnons\cite{DONIACH}. The substantial reduction in the Wilson Ratio for $x$ = 1 shown in Fig.~\ref{WILSONFIG}(inset) could be evidence in favour of electron-phonon enhancement.  Although electron-phonon interactions are thought to be relatively weak in these materials\cite{BOERI}, we should remember that any electron interaction process that is strongly volume dependent will increase the electron-phonon coupling. The peak in $\gamma_n$ at 15K for $x$ = 1 is  reduced by less than 0.15$\%$ in magnetic fields up to 13T (for which the Zeeman energy  $\mu_BH/k_B$ = 9K). This could be a problem for theories involving spin fluctuations and might point towards more general correlations of the Fe 3d electrons as mentioned in Ref.~\onlinecite{TERASHIMA}.

\subsection{Superconducting State.}
\begin{figure}
\centering
\includegraphics[width=\linewidth]{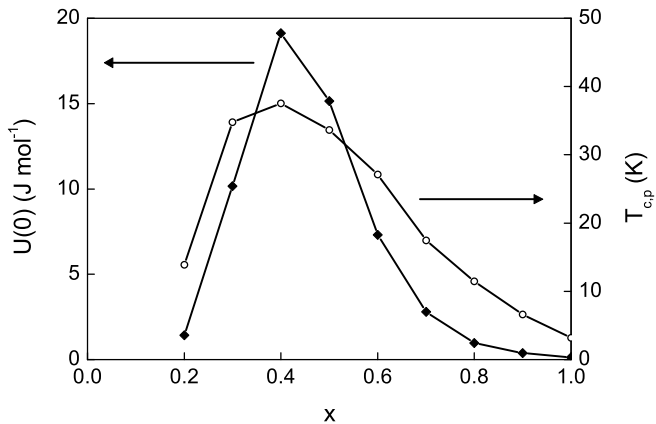}%
\caption{
Doping dependence of the superconducting condensation energy and the main-peak transition temperature $T_{c,p}$.
} 
\label{U0FIG}
\end{figure}

\begin{figure}
\centering
\includegraphics[width=\linewidth]{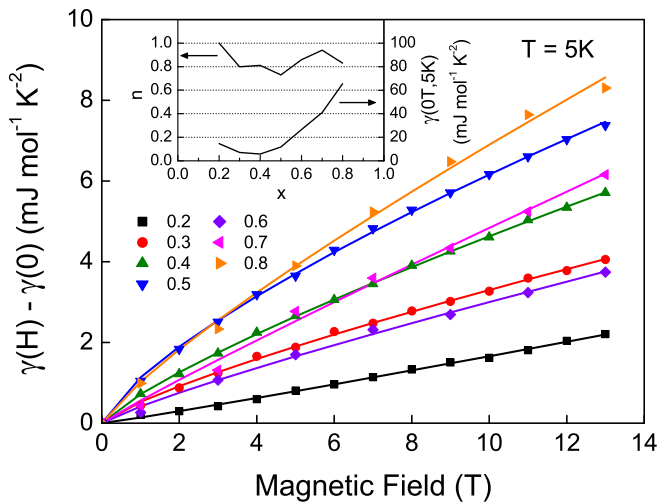}%
\caption{
(Color online) The change in electronic specific heat at 5K as a function of magnetic field for $x$ = 0.2 to 0.8. Lines show fits to $H^n$ with values of $n$ shown in the the inset. In the clean limit, $n$ = 1 (0.5) is expected for a single $s$-wave ($d$-wave) gap. Also shown in the inset is the doping dependence of the zero-field electronic specific heat at 5K, which for $x>0.4$ is significantly higher than its limiting value at low temperatures.
} 
\label{DGH5KFIG}
\end{figure}

Several important properties of the superconducting condensate can be determined directly from the specific heat data shown in Figs.~\ref{GHVSTFIG} and \ref{ENTROPYFIG}. The superconducting condensation energy $U(0)$ is the free energy difference at $T$ = 0 between normal and superconducting states in zero field, and is given by $U(0)=\int_0^{T_c}{(S_n-S_s)dT}$. Values for $U(0)$ are shown in Fig.~\ref{U0FIG} and peak sharply at $x$ = 0.4. The free energy difference between 0 and $H$, $\Delta F(H,T) =\int_0^T{[S(H)-S(0)]dT}$ yields the magnetisation $M=-(\partial\Delta F/\partial H)_T$ in the mixed state and from that the superfluid density and critical fields. This will be the subject of a further publication. Finally, the field dependence of $\gamma(H,T)$ in the mixed state at low-$T$ can be used to determine the pairing symmetry. In the clean limit a single $s$-wave gap gives rise to a linear $H$-dependence\cite{LINEARH}, while a $d$-wave gap results in a $\sqrt{H}$-dependence\cite{VOLOVIK}.
$\gamma(H)-\gamma(0)$ at 5K is shown in Fig.~\ref{DGH5KFIG} for $x$ = 0.2 to 0.8. The data is well described by a slightly sub-linear $H^n$ power law with $n$ ranging from 0.75 to 1.0 (see the inset to Fig.~\ref{DGH5KFIG}). 
Sublinear behaviour has been interpreted terms of a multiband $s\pm$-wave state, comprising either two unequally sized isotropic $s$-wave gaps in the presence of impurity scattering, as proposed by Bang\cite{BANG}, or an isotropic gap in combination with an anisotropic gap, as proposed by Wang\cite{WANG2}. We see from the inset to Fig.~\ref{DGH5KFIG} that $\gamma(0,T=5\mathrm{K})$ is much smaller for $x$ = 0.3 to 0.5 where $n\approx0.75$. It is quite possible  that all samples would show sub-linear $H$-dependence  in the low-$T$ limit. A detailed analysis taking into account the three bands described below and their $T$ dependence would be needed to obtain meaningful values of $n$.

\begin{figure*}
\centering
\includegraphics[width=\linewidth]{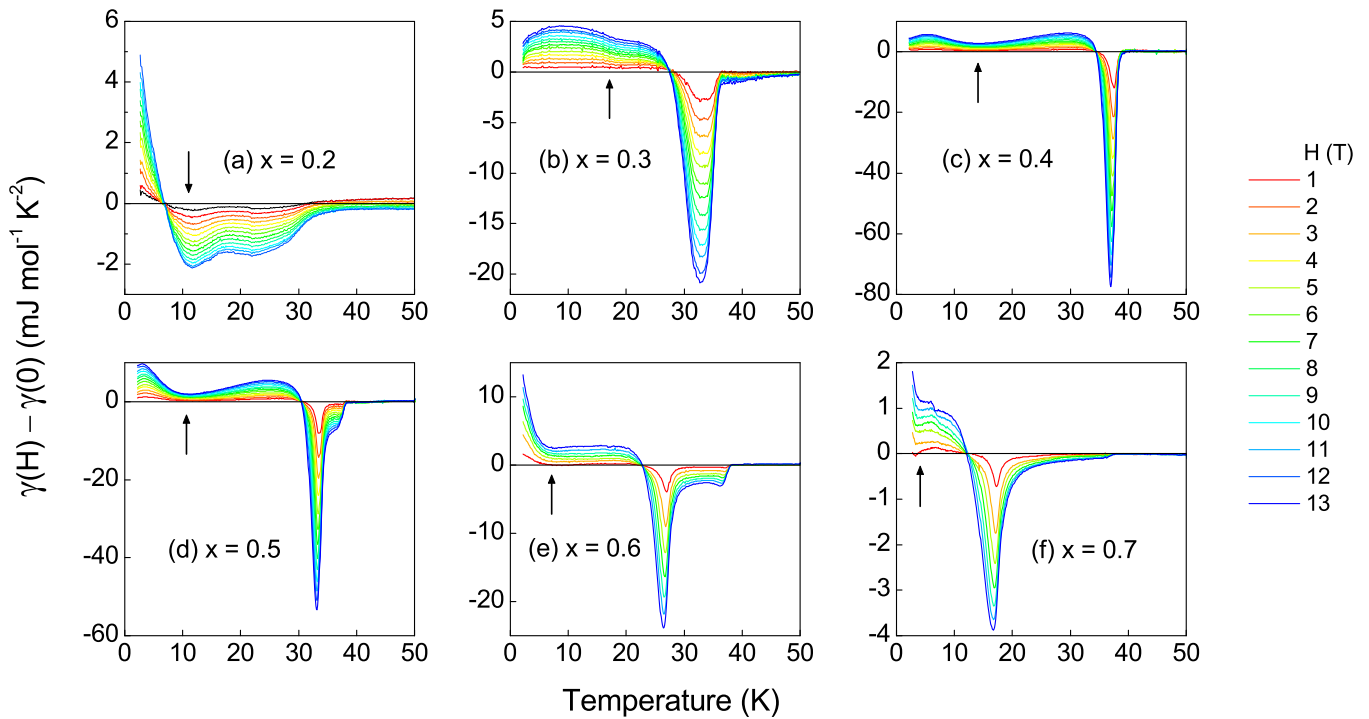}%
\caption{
(Color online) (a) to (f) The magnetic field dependent change in electronic specific heat vs temperature up to 13T for $x$ = 0.2 to 0.7. Arrows mark the temperature of the knee feature, $T_k$.
} 
\label{DGHTFIG}
\end{figure*}

\begin{figure}
\centering
\includegraphics[width=\linewidth]{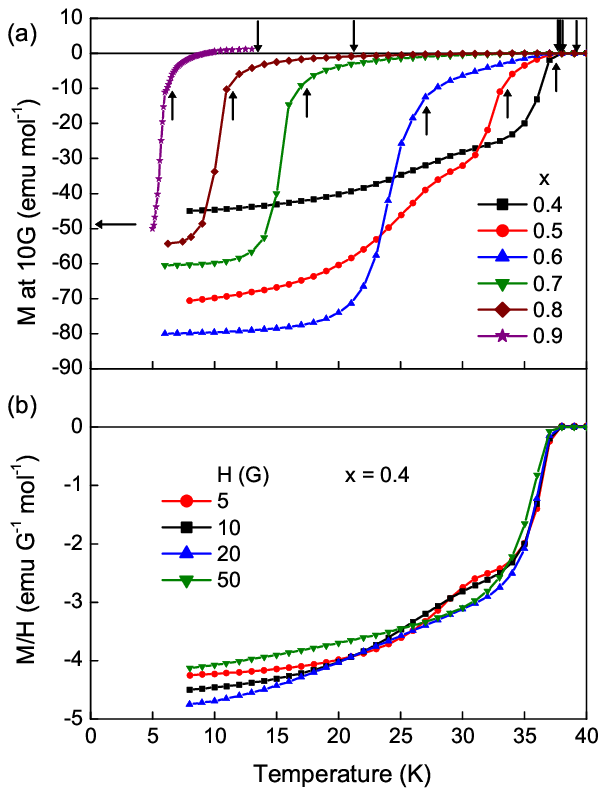}%
\caption{
(Color online) (a) Zero-field-cooled magnetization measured in a 10 Gauss magnetic field for $x$ = 0.4 to 0.8. Upward and downward arrows denote the main peak temperature ($T_{c,p}$) and shoulder onset temperature ($T_{c,s}$) respectively. (b) Zero-field-cooled magnetization (normalised by magnetic field) for $x$ = 0.4 in magnetic fields of 5 to 50 Gauss. The structure between 25 and 35K is due to intergrain coupling.
} 
\label{ZFCFIG}
\end{figure}

\begin{figure}
\centering
\includegraphics[width=\linewidth]{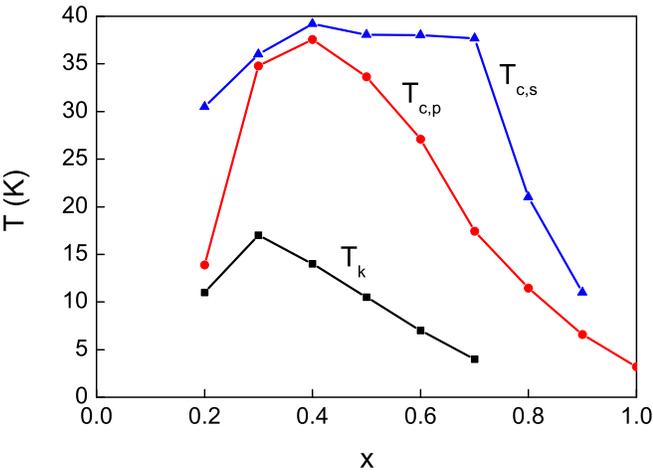}%
\caption{
(Color online) Temperatures $T_{k}$, $T_{c,p}$ and $T_{c,s}$ associated with the knee, peak and shoulder features respectively.
}
\label{TKPSFIG}
\end{figure}

Like $\gamma_n$, the superconducting-state $\gamma$ displays a rich progression with doping. This can be clearly seen from the magnetic field dependence $\gamma(H,T)$ for $x$ = 0.2 to 1 shown in Fig.~\ref{GHVSTFIG} and from the change with field $\Delta\gamma(H,T)=\gamma(H,T)-\gamma(0,T)$ shown in Fig.~\ref{DGHTFIG}. Note that the phonon contribution to the raw data is independent of field and does not contribute to $\Delta\gamma(H,T)$. In contrast to a previous study\cite{TANAKAY} we observe superconducting anomalies in all samples from $x$ = 0.2 to 1.0. For most of the superconducting compositions we can identify three distinct features in plots of $\gamma(H,T)$ (Fig.~\ref{GHVSTFIG}) and $\Delta\gamma(H,T)$ in Fig.~\ref{DGHTFIG}  which appear to correspond to different SC gaps in three bands. The temperatures associated with these features are shown in Fig.~\ref{TKPSFIG}. 
The most obvious feature  is a relatively sharp mean-field-like peak at $T_p$ reflecting the collapse of a gap with $T$ = 0 magnitude $\Delta_p(0)/k_B=\alpha_pT_{c,p}$, where $T_{c,p}$ is taken to be the temperature of maximum negative slope of $\gamma(T)$ (slightly above the peak temperature $T_p$). This feature appears as a sharp negative peak at $T_p$ in $\Delta\gamma(H,T)$ in Fig.~\ref{DGHTFIG}. 
At lower temperature there is a broad Schottky-like anomaly (the ``knee'') peaking at $T_k$. The progressive suppression of the knee in the vicinity of $T_k$ with increasing magnetic field, clearly seen in Fig.~\ref{DGHTFIG}, and the rapid increase in $\gamma(H,T)$ at lower temperatures seen in Fig.~\ref{GHVSTFIG} confirm that the ``knee'' is of superconducting origin and is not an artifact of errors in our phonon correction. Since the peak is broad, the underlying SC gap must be essentially $T$-independent in the region $T\sim T_k$, with approximate magnitude $\Delta_k(0)/k_B \sim 2.2T_k$. The onset temperature of the knee gap is uncertain but is at least as high as $T_p$ since no further anomaly is seen between $T_k$ and $T_p$. 
The highest temperature feature clearly visible in the zero field data in Fig.~\ref{GHVSTFIG} for $x$ = 0.5, 0.6 and 0.7, and in the onset of an $H$-dependent suppression for $x$ = 0.8 and 0.9, is a broad shoulder extending above $T_p$ with an abrupt onset signifying a phase transition at $T_{c,s}$. $T_{c,s}$ is close to $T_m\approx38$K for $x$ = 0.5 to 0.7 and $\approx$ 21K for $x$ = 0.8 and 11K for $x$ = 0.9. We note that $T_{c,s}$ also coincides with the onset of diamagnetism (see Fig.~\ref{ZFCFIG}) and is the true superconducting transition temperature. Examination of Fig.~\ref{GHVSTFIG} shows that field dependent shifts are comparable in magnitude for both the main peak and shoulder for $x$ = 0.5 to 0.8. This implies that values for the upper critical field will also be comparable for both features. Finally as shown in Fig.~\ref{GHVSTFIG} , for dopings, $x$ = 0.3 to 0.6 where the ``knee'' gap is large enough to estimate the initial $T$-dependence of $\gamma(T)$, we find a small but finite $\gamma(T=0)$.  Even in samples where $\gamma$ is relatively large at our lowest temperature a small value of $\gamma(0)\ll\gamma_n$ can be inferred from plots of $S/T$ in Fig.~\ref{ENTROPYFIG}. A small residual $\gamma(0)$ has been found for $x$ = 1 in measurements down to 0.1K\cite{KIM} confirming our conclusions from $S/T$ (Fig.~\ref{ENTROPYFIG}). These values of $\gamma(0)$ are summarised later in Fig.~\ref{ALPHAFIG}(c). The conclusion that $\gamma(0)$ is small for all $x$ is important since it confirms the existence of a low temperature knee in $\gamma(T)$ for all of the higher values of $x$, and also demonstrates that the small magnitude of the anomalies at $T_p$ for high $x$ is not due to non-superconducting regions in the sample or to strong pair breaking. The small residual $\gamma(0)$ may result from pair breaking in one or more of the gaps, a low level of impurities, or an additional non-superconducting band with a very small DOS. 
For $x$ = 0.2 the strong suppression of the superconducting transition in the magnetically ordered phase makes identification of the three features discussed above far less clear. The structure in $\Delta\gamma(H,T)$ in Fig.~\ref{DGHTFIG}(a) reveals that the rather featureless broad superconducting anomaly shown in Fig.~\ref{GHVSTFIG}(a) is in fact composed of two peaks at 11K and 23K, which may perhaps be attributed to the main-peak and shoulder bands respectively. The weak negative curvature in zero field and rapid increase with $H$ below the crossing point at 7K may suggest a small superconducting gap possibly associated with the ``knee'' band. 

\subsubsection{Knee.}
As shown in Figs.~\ref{GHVSTFIG}, \ref{DGHTFIG} and~\ref{TKPSFIG} the peak temperature $T_k$ of the broad ``knee'' feature decreases with $x$ from around 17K for $x$ = 0.3 to 3K for $x$ = 0.7 and in fact is still present at $\sim$ 0.7K\cite{KIM} for $x$ = 1. However, the amplitude of the knee grows with $x$ due to the increasing dominance of the normal state DOS for this band (Sec.~\ref{secBANDSTRUC}), and at low-$T$ and high-$x$ this feature makes the largest contribution to $\gamma(T)$.

\subsubsection{Main Peak.}
\begin{figure}
\centering
\includegraphics[width=\linewidth]{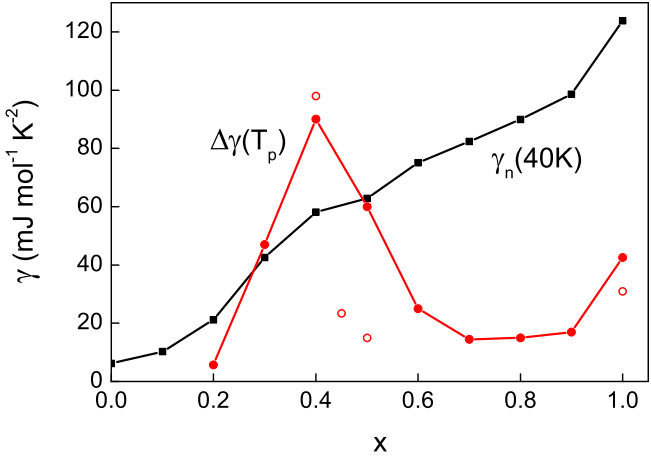}%
\caption{
(Color online) Doping dependence of the jump $\Delta\gamma(T_p)$ in the electronic specific heat ($\Delta\gamma$) at $T_p$, compared with published values for single crystal samples (open circles)\cite{MU,DONG,NI,KIM}. Also shown is the normal-state electronic specific heat at 40K.
} 
\label{DGFIG}
\end{figure}

Transition temperatures for Ba$_{1-x}$K$_x$Fe$_2$As$_2$ quoted in the literature\cite{JOHRENDT,AVCI2} correspond most closely to the transition temperatures $T_{c,p}$ of the mean-field-like peaks shown in Fig.~\ref{TKPSFIG}. For $0.3\leq x\leq0.7$, $T_{c,p}(x)$ is approximately parabolic with a maximum value $T_m$ = 38.5K at $x_m$ = 0.39. However, evidence presented below suggests that for $x$ = 0.4 the main peak and shoulder anomalies may almost coincide, and that $T_{c,p}$ and therefore also $T_m$ may be $\sim$ 1K lower than the values quoted above. At higher $x$, $T_{c,p}$ decreases more slowly and for $x$ = 1 we find $T_{c,p}$ = 3.2K. With the disappearance of the magneto/structural transition just below $x$ = 0.3, the anomaly height $\Delta\gamma(T_{c,p})$ at $T_p$ increases rapidly to a maximum at $x$ = 0.4, as shown in Fig.~\ref{DGFIG} . For $x$ = 0.4 and 0.5 the anomaly heights are comparable with or larger than published single crystal data\cite{MU,DONG}. The decrease in $\Delta\gamma(T_{c,p})$ by $30\%$ in $x$ = 0.5 and a further factor of two for $x$ = 0.6 coincides with the growth of the ``shoulder'' above $T_p$. $\Delta\gamma(T_{c,p})$ is relatively constant between $x$ = 0.6 and 0.9 where we see evidence for a ``shoulder'', but increases for $x$ = 1 where no evidence for a ``shoulder'' is observed (Fig.~\ref{DGFIG}). Since the residual $\gamma(0)$ is small and $\gamma_n$ increases continuously across the series (Fig.~\ref{DGFIG}), an increase in $\Delta\gamma(T_{c,p})$ for $x>0.4$ would be expected on a single band scenario. In a multi-band situation however, $\Delta\gamma(T_{c,p})$ would be roughly constant if the contributions to $\gamma_n$ from  the bands with larger gaps are approximately doping independent, as suggested by explicit fits to the data in Section \ref{secBANDSTRUC} (Fig.~\ref{ALPHAFIG}).  So the very sharp fall in $\Delta\gamma(T_{c,p})$ is unexpected, and seems to result more from the growth of the shoulder.

The field dependence of the main peaks shown in Figs.~\ref{GHVSTFIG} and \ref{DGHTFIG} provide clear evidence for short coherence lengths and low dimensionality. Firstly the peaks broaden and reduce in height with an almost $H$-independent onset. This behaviour is typical of short coherence length superconductors such as the high-$T_c$ cuprates and is in sharp contrast to the progressive shift to lower temperature without change in shape observed in classical superconductors\cite{JUNODREVIEW}. Secondly for $x$ = 0.2 to $0.8$ there is a well defined crossing point in $\gamma(H,T)$ 5 to 10K below $T_p$ over a wide range of fields which is also observed in the highly anisotropic cuprate Bi$_2$Sr$_2$CaCu$_2$O$_8$\cite{JUNOD}. In zero field, superconducting fluctuations invariably extend well above the temperature at which an $H$-dependence is first observed, and can in the present system be easily distinguished from the shoulder by their positive curvature and absence of an onset temperature. Zero-field fluctuations above $T_{c,p}$ can be seen in the data for $\gamma(H,T)$ in Fig.~\ref{GHVSTFIG} and in $d\gamma_{sn}/dT$ shown in Fig.~\ref{DGDTFIG}, and in all cases appear to diverge towards $T_{c,p}$ and never towards $T_{c,s}$. 
This term can be well fitted both near $T_{c,p}$ and above $T_{c,s}$ by an expression for 3D-2D Gaussian fluctuations\cite{LORAMFLUCS}, 
\begin{equation}
\gamma^{fluc} =  \frac{B_{fluc}}{T\sqrt{\tau^2+\tau \tau^*}}
\label{eq:FLUC}
\end{equation} 
where $B_{fluc} = (R/4\pi)(a/\xi_{ab})^2$,  $\tau = |T/T_{c,fl}-1|$ and $\tau^* = (\xi_c/c)^2$ is the 3D-2D crossover temperature. $R$ is the universal gas constant, $a$ and $c$ are the lattice parameters from Fig.~\ref{LATTICEFIG}, and $\xi_{ab}$ and $\xi_c$ are the $ab$-plane and $c$-axis superconducting coherence lengths at $T$ = 0. Values for all the parameters for $H$ = 0 are shown in Table.~\ref{FLUCTABLE}. 
For most of the samples, good fits to the 3D-2D Gaussian fluctuation expression were obtained taking $T_{c,fl}=T_{c,p}$, the main peak transition temperature. Deviations from the fit are only visible close to $T_{c,p}$, as demonstrated by the abrupt downturns in the corrected curves in Figs.~\ref{GSNFIG}(a) and \ref{GSNFIG}(b). However for $x$ = 0.4 where no shoulder is observed, there is an abrupt change of the fluctuation $T$-dependence close to 38K which is similar to that seen in samples with a shoulder. Because of this abrupt change a good fit to the Gaussian fluctuation expression above 38K could only be obtained if $T_{c,fl}$ was taken to be $\sim$ 1K lower than $T_{c,p}$. This may be indirect evidence that the main peak and shoulder anomalies are almost coincident for this sample and the true $T_{c,p}$ for $x$ = 0.4 is $\sim$ 1K lower than the value 37.55K quoted in Table~\ref{FLUCTABLE}.  
Values of $B_{fluc}$ are reasonably reliable but values of $\tau^*$ are not very reliable for samples with a shoulder. For $x$ = 0.3, $\tau^*$ is very sensitive to the value taken for $T_{c,fl}$ which is difficult to estimate since the main peak and shoulder are difficult to distinguish. Values for the coherence length $\xi_{ab} \sim 20$\AA\ that we find from the fluctuation term in BKFA are comparable with those found in the cuprates and are consistent with the large values of upper critical field $H_{c2}$ that can be inferred from plots of $\gamma(H,T)$ in Fig.~\ref{GHVSTFIG}.

\begin{table*}
\begin{tabular}{|p{1cm}|p{1cm}|p{1cm}|c|p{1cm}|p{1cm}|p{1cm}|p{1cm}|c|p{1cm}|}
\hline
$x$ & $T_{c,p}$ & $T_{c,fl}$  & $B_{fluc}$ & $\tau^*$ & $\xi_{ab}$ & $\xi_{c}$ & $T_{c,s}$ & $A$ & $m$\\
& (K) & (K) & (mJ/mol K$^2$) & & (\AA) & (\AA) & (K) & (mJ/mol K$^2$) & \\
\hline
0.3 & 35.88 & 34.1 & 25 & 0.02 & 20.0 & 1.0 & & & \\
0.4 & 37.55 & 36.5 & 43 & 0.19 & 15.3 & 2.9 & & & \\
0.5 & 33.75 & 33.75 & 28 & 0.06 & 19.0 & 1.7 & 37.63 & 49 & 0.5\\
0.6 & 27.12 & 27.12 & 21 & 0.12 & 21.9 & 2.3 & 37.53 & 25 & 0.5\\
0.7 & 17.5 & 17.5 & 18 & 0.14 & 23.6 & 2.5 & 37.51 & 9.45 & 0.81\\
\hline
\end{tabular}
\caption{Parameters extracted from 2D-3D Gaussian fluctuation fits.}
\label{FLUCTABLE}
\end{table*} 

\subsubsection{Shoulder.}
The onset of the shoulder at $T_{c,s}$ is abrupt, signifying a phase transition, and is the temperature at which diamagnetism is first observed (Fig.~\ref{ZFCFIG}), confirming its superconducting origin. Fluctuations diverging towards $T_{c,s}$ are at least two orders of magnitude smaller than fluctuations diverging towards $T_{c,p}$. For $x$ = 0.5, 0.6 and 0.7, $T_{c,s}\sim$ 38K is close to $T_m$, the maximum value of $T_c$ (Fig.~\ref{TKPSFIG}). It then falls to $\sim$ 21K and 11K for $x$ = 0.8 and 0.9. 

\begin{figure}
\centering
\includegraphics[width=\columnwidth]{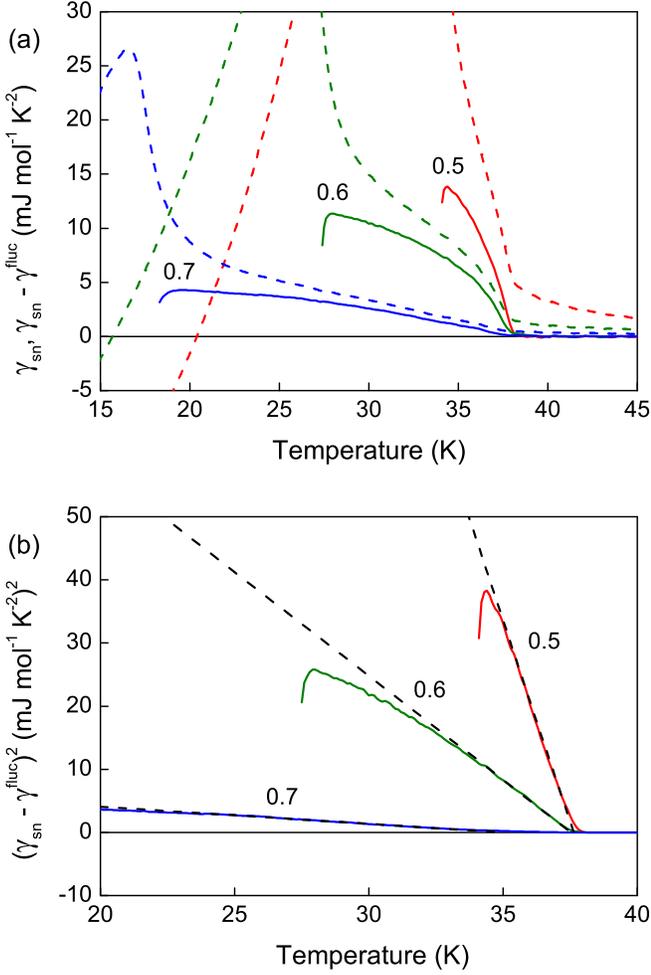}
\caption{(Color online) (a) The shoulder feature in $\gamma_{sn}=\gamma_s-\gamma_n$ (dotted lines) and $\gamma_{sn}-\gamma^{fluc}$ (solid lines) for $x$ = 0.5, 0.6 and 0.7. (b) ($\gamma_{sn}-\gamma^{fluc})^2$ vs $T$ illustrating the initial $\sqrt{1-T/T_{c,s}}$ temperature dependence for $x$ = 0.5 and 0.6. For $x$ = 0.7 see text.
}
\label{GSNFIG}
\end{figure}

The shoulder $T$-dependence is unique as far as we are aware, and for $x$ = 0.5 to 0.7 is well described over a substantial range below $T_{c,s}$ by the expression
\begin{equation}
\gamma_{sn} - \gamma_{fluc} = A(1-T/T_{c,s})^m	
\label{eq:gfluc}
\end{equation} 
with values of $A$, $T_{c,s}$ and $m$ shown in Table~\ref{FLUCTABLE}. The amplitude $A$ of the shoulder decreases rapidly as $T_{c,s} - T_{c,p}$ increases and for $x$ = 0.5 and 0.6 the exponent $m$ is close to 0.5, as demonstrated by the linearity of $(\gamma_{sn}-\gamma_{fluc})^2$ vs $T$ shown in Fig.~\ref{GSNFIG}(b). For $x$ = 0.7 we find an exponent $m\sim$ 0.8. For this sample the rather large $T_{c,s} - T_{c,p}\sim$ 20K and uncertainty in the $T$-dependences of the normal state $\gamma_n$ and the fluctuation term complicate the determination of the magnitudes and $T$-dependences of the shoulder. For $x$ = 0.8 and 0.9 there is a more or less abrupt onset at $T_{c,s}$ but the shoulder has a positive curvature corresponding to $m>1$. For convenience we will refer below to Eq.~\ref{eq:gfluc} as the ``$A\sqrt{1-T/T_{c,s}}$'' dependence of the shoulder recognising that this $T$-dependence is only correct for the $x$ = 0.5 and 0.6 samples. 
	We consider four possible causes for the shoulder between $T_p$ and $T_s$, a) That $T_{c,s}$ is the transition temperature for the ``knee'' gap; b) inhomogeneity in the local doping $x$ giving rise to a spread of local $T_c$'s and a consequent broadened anomaly; c) intrinsic electronic inhomogeneity; d) a third band with energy gap $\Delta_s$ which opens at $T_s$. We explore these possibilities below.

\textbf{a)} Attributing the shoulder to the onset of the ``knee'' gap is tempting due to its simplicity. However, since $T_{c,s}\gg\Delta_k$, the entropy difference $S_{ns}$ above $T_p$ is too small to account for the weight in the shoulder and would result at most in only a very weak anomaly at $T_s$. We therefore reject this hypothesis.

\begin{figure}
\centering
\includegraphics[width=\linewidth]{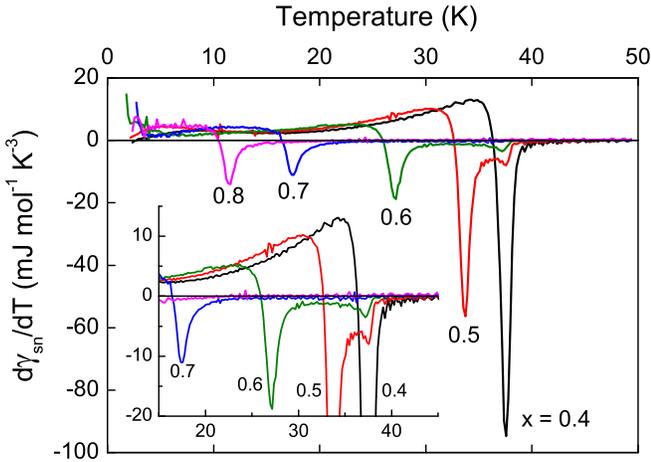}%
\caption{
(Color online) Temperature derivative of $\gamma_{sn}=\gamma_s-\gamma_n$ for $x$ = 0.4 to 0.8. The inset shows an expanded plot of the 15 to 45K region.
}
\label{DGDTFIG}
\end{figure}

\textbf{b)} In principle the shoulder could be caused by inhomogeneity in local doping $x$ leading to a distribution of $T_c$'s. In the present case where $T_c(x)$ has a maximum at  $T_m$, the shoulder would still have a sharp onset at $T_m$ as observed. However considerable insight into the question of chemical inhomogeneity is given by the plots of $d\gamma_{sn}/dT$ in zero field shown in Fig.~\ref{DGDTFIG} for $x$ = 0.4 to 0.8. For all samples we see a sharp negative peak in $d\gamma_{sn}/dT$ resulting from the mean-field like anomaly at $T_p$ and, for the 0.5 and 0.6  samples, a broad feature extending up to $T_{c,s}$ due to the shoulder. Since these two features can be clearly distinguished we will consider them separately. 
 
\begin{figure}
\centering
\includegraphics[width=\linewidth]{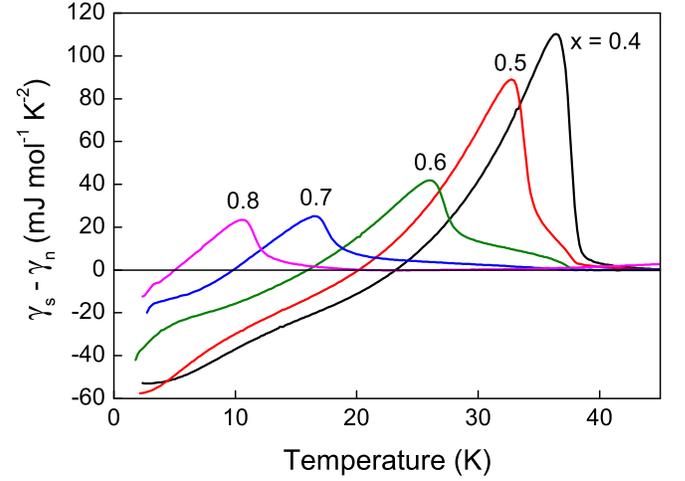}%
\caption{
(Color online) Electronic specific heat with the normal-state part subtracted for $x$ = 0.4 to 0.8.
} 
\label{GSNVSTFIG}
\end{figure}

The sharp negative peak can be fitted to a Gaussian function $d\gamma_{sn}/dT=\Delta\gamma_p/(\sqrt{2\pi}\sigma_T)\exp(- (T-T_c)^2/2\sigma_T^2)$, where $\Delta\gamma_p$ is the ideal entropy conserving step height in $\gamma_{sn}$ in the absence of broadening and $\sigma_T$ is the standard deviation. Values of $T_c$, $\Delta\gamma_p$ and $\sigma_T$ are given in Table~\ref{SPREADTABLE}. If we assume that the spread in $T_c$ for the sharp peak is due to a spread in local doping $x$, then we can use the $T_{c,p}(x)$ data in Fig.~\ref{U0FIG} to convert them into the standard deviation in $ x$, $\sigma_x$.  As shown in Table~\ref{SPREADTABLE}, these are remarkably small. Furthermore it is evident from Figs.~\ref{GHVSTFIG} and \ref{GSNVSTFIG} that the shoulder and main peak make comparable contributions to the total height of the anomaly. Interpreted in terms of inhomogeneous doping this requires a bi-modal distribution with around half of the sample being very close to the nominal composition and half having a broad range of $x$. We have made a detailed analysis of this situation using arguments summarised in the Appendix starting from the formula:
\begin{equation}
\gamma_{sn}(T)=\int{\gamma_{sn}(T/T_c)P(T_c)dT_c}	
\label{eq:L1}
\end{equation}
where $\gamma_{sn}=\gamma_s-\gamma_n$ and $P(T_c)$ is the normalised probability distribution of local $T_c$'s. The function $\gamma_{sn}(T/T_c)$ can either represent an unbroadened mean-field transition with a discontinuous jump $\Delta\gamma$ at $T_c$ and is zero for $T>T_c$ or the actual experimental data for $x$ = 0.4 where no shoulder is observed. In both cases we conclude that arguments involving chemical inhomogeneity are contradicted by the linear $x$-dependence and symmetric broadening seen in the X-ray spectra (Figs.~\ref{LATTICEFIG}(a)$-$(c)).

\begin{table}
\begin{tabular}{|p{1cm}|p{1cm}|c|p{1cm}|p{1cm}|p{1cm}|}
\hline
$x$ & $T_{c,p}$ & $\Delta\gamma_p$  & $\sigma_T$ & $\sigma_x$  \\
& (K) & (mJ/mol K$^2$) & (K) & \\
\hline
0.4 & 37.55 & 90 & 0.41 & 0.047 \\
0.5 & 33.75 & 60 & 0.42 & 0.008  \\
0.6 & 27.12 & 25 & 0.54 & 0.008  \\
0.7 & 17.50 & 14.5 & 0.52 & 0.005  \\
0.8 & 11.46 & 15 & 0.43 & 0.007 \\
0.9 & 6.6 & 17 & 0.40 & 0.011 \\
1.0 & 3.2 & 42.5 & 0.18 & 0.005 \\
\hline
\end{tabular}
\caption{Transition temperature ($T_{c,p}$), ideal step height ($\Delta\gamma_p$) and standard deviations in transition temperature ($\sigma_T$) and doping ($\sigma_x$) of the main peak feature.}
\label{SPREADTABLE}
\end{table} 

A further puzzling feature is the absence of a significant fluctuation term diverging at or near the onset of the shoulder at $T_{c,s}$. Integrating Eq.~\ref{eq:FLUC} for $\gamma^{fluc}$ over the distribution $P(T_c)\propto 1/\sqrt{1-T_c/T_{c,s}}$ expected for inhomogeneity extending through $x_m$ gives a fluctuation term $\gamma^{fluc}\propto 1/\sqrt{T-T_{c,s}}$ that diverges at $T_{c,s}$. If $B^{fluc}$ is comparable with the values found for the main peak, this term would be very much larger than that observed, and its absence is further evidence against an explanation for the shoulder in terms of spatial inhomogeneity.

An explanation for the shoulder in terms of doping inhomogeneity therefore faces severe challenges. i)  How can a very sharp main peak at $T_p$ with small $\sigma_x\sim$ 0.006 (Table~\ref{SPREADTABLE}) coexist with a shoulder with a much larger spread $\sim2|x_0-x_m|\sim$ 0.2 to 0.6, where $x_0$ is the nominal concentration? (See Appendix). ii) The large weight $\sim50\%$ in the shoulder with a substantial probability $P(x_m)$ at $x_m$ should be clearly visible in X-ray spectra. In fact the X-ray spectra shown in Fig.~\ref{LATTICEFIG} are relatively sharp with an $x$-independent half width and no evidence for a shoulder towards or beyond $x_m$. iii) If there is doping inhomogeneity we would expect this to be symmetrical about the nominal doping $x_0$ and the average to be close to $x_0$, contrary to the conclusions in the Appendix. iv) This interpretation cannot explain the divergence of the amplitude $A$ of the shoulder as $T_{c,s}-T_{c,p}$ decreases, or the almost complete absence of fluctuations diverging at $T_{c,s}$.

For all these reasons we believe the samples are spatially homogeneous with an root-mean-square (rms) width $\sim$0.006 given by that of the main peak at $T_p$. This is in fact an upper limit as the width will also contain a contribution from instrumental broadening of the transition. We conclude that it is extremely unlikely that the shoulder is due to local doping inhomogeneity.

\textbf{c)} Intrinsic electronic inhomogeneity.
Even without chemical inhomogeneity, local electronic inhomogeneity may give rise to a distribution of gaps with onset temperatures from $T_{c,p}$ to $T_{c,s}$. Such a situation might arise due to local variations in the Fe-As-Fe bond angle, the value of which can have a strong influence on the presence or absence of certain sheets of the Fermi surface\cite{USUI}. Frustration in the sign of the superconducting gaps in a 3-band $s\pm$ scenario may also lead to a local spread of gap magnitudes. Any band or gap unaffected by these effects would contribute to the sharp anomaly. These local effects on bands and superconducting gaps would not affect the X-ray spectra and therefore this interpretation would not be subject to many of the objections raised above against chemical inhomogeneity. However, as seen above, the anomalous $\sqrt{1-T/T_m}$ dependence of $\gamma(T)$ in the shoulder region requires a probability distribution $P(T_c)\sim 1/\sqrt{1-T_c/T_m}$ that diverges at $T_m$. This is expected for chemical inhomogeneity extending through $x_m$ with a parabolic $T_c(x)$, but would be harder to explain for intrinsic electronic inhomogeneity. 

\textbf{d)}   Our last hypothesis is that the shoulder results from a third band with a superconducting gap $\Delta_s$ with an unconventional $T$-dependence near $T_{c,s}$. Evidence for three gaps has been observed in electron-doped BaFe$_{1.87}$Co$_{0.13}$As$_2$\cite{KIMKW} and multiple bands are found to cross the Fermi level in BKFA\cite{KORDYUK}, so the prospect of multiple gaps is not unreasonable. 

The complex transitions seen here in BKFA appear to suggest a minimum of three bands and three superconducting gaps. Theories of two coupled gaps\cite{SUHL,KOGAN} invariably predict a sharp peak in $\gamma(T)$ at the initial onset temperature $T_c$, and a broad Schottky like anomaly at lower temperatures. This agrees with the behaviour seen in the well established two gap material MgB$_2$\cite{BOUQUET}, and accounts for the ``knee'' and ``main peak'' features in our data. Two-gap theories cannot however explain the additional ``shoulder''  feature in BKFA and most importantly, the absence of a jump in $\gamma(T)$ at the onset of long range order at $T_{c,s}$.

The $T$-dependence of the gap close to $T_{c,s}$ can be determined from the $\gamma(T)\sim(T_{c,s}-T)^{0.5}$ $T$-dependence in the shoulder region. If the quasi-particle energies have the BCS dependence
\begin{equation}
E=\sqrt{\epsilon^2+\Delta(T)^2}
\label{eq:E}
\end{equation}
it is easily shown that close to the transition when $\Delta(T)\ll k_BT$, the entropy $S_{ns}=S_n-S_s$ is given by 
\begin{equation}
TS_{ns} = N(0)\Delta(T)^2
\label{eq:TS}
\end{equation}
for	$\Delta(T)\ll k_BT$,
where $N(0)=3/(2\pi^2k_B^2)\gamma_n$ is the renormalised normal state DOS/spin at $E_F$ and $\gamma_n\equiv\gamma_n(T=0)$. If, as in a usual mean field transition, $\gamma_{ns}$ has a step $\Delta\gamma(T_c)$ at $T_c$ then $S_{ns}=\Delta\gamma(T_c)\cdot(T_c-T)$ just below $T_c$ and Eq.~\ref{eq:TS} gives the expected mean-field dependence $\Delta(T)\sim(T_c-T)^{0.5}$. If there is no step at $T_c$ but instead $\gamma_{ns}\sim(T_c-T)^{0.5}$ as we observe for the shoulder, then $S_{ns}\sim(T_c-T)^{1.5}$ and from Eq.~\ref{eq:TS}, $\Delta(T)\sim(T_c-T)^{0.75}$. We are unaware of any theoretical treatment that predicts this limiting $T$-dependence. Note that for a $k$-dependent gap or for multiple gaps ($\Delta_i$), $\Delta^2(T)$ in the above expressions should be replaced by a Fermi surface average $\Delta_{rms}^2(T) =\sum{g_i\Delta_i^2}$ and $g_i=\gamma_{n,i}/\gamma_n$.

Ferrell\cite{FERRELL} has derived the following exact expression for $\Delta^2(T)$ for a weak-coupling $s$-wave superconductor in terms of the superconducting free energy $F_{ns}$ and entropy $S_{ns}$ which is valid over the entire range $0<T<T_c$
\begin{equation}
2F_{ns}(T)+TS_{ns}(T)=N(0)\Delta^2(T) 
\label{eq:L6}
\end{equation} 	
where $F_{ns}= -\int_0^{T_c}{S_{ns}dT}$. This result is also valid for an anisotropic gap if $\Delta^2(T)$ is replaced by a Fermi surface average $\Delta_{rms}^2(T)$, and we will assume that it is approximately correct for coupled multiple bands. This gives the standard BCS expression at $T$ = 0, $\Delta_{rms}^2= 2U(0)/N(0)$, where $U(0)=F_{ns}(T=0)$ is the SC condensation energy, and also the expression $\Delta_{rms}^2(T\sim T_c)=(2\pi^2k_B^2/3)TS_{ns}/\gamma_n=TS_{ns}/N(0)$ near $T_c$ where $\Delta_{rms}/k_B\ll T$. We will therefore assume that for coupled gaps Eq.~\ref{eq:L6} gives a good approximation for $\Delta_{rms}^2(T)$ over the entire temperature range, and that the temperature derivative of Eq.~\ref{eq:L6} 
\begin{equation}
C_{ns}-S_{ns}=T^2\frac{d(S_{ns}/T)}{dT} = N(0)\frac{d\Delta_{rms}^2}{dT}=\frac{3\gamma_n}{2\pi^2k_B^2}\frac{d\Delta_{rms}^2}{dT}
\label{eq:L7}
\end{equation}
gives a good approximation to the slope $d(\Delta_{rms}^2)/dT$, (ignoring the $T$-dependence of $\gamma_n$). We note that for strong coupling, $N(0)$ in Eqs.~\ref{eq:TS}-\ref{eq:L7} is smaller than its normal state value because of the effect of the superconducting gap on the renormalisation\cite{SCALAPINO}. Thus for strong coupling superconductors using Eqs.~\ref{eq:TS}-\ref{eq:L7} with the normal state value for $N(0)$ underestimates the true values for $\Delta_{rms}^2$ and $d(\Delta_{rms}^2)/dT$. For example, for the strong coupling superconductor Pb, $\Delta(0)$ deduced from $U(0)$ is $\sim12\%$ lower than the gap $\Delta(0)$ found from tunnelling experiments\cite{SCALAPINO}. 

\begin{figure}
\centering
\includegraphics[width=7cm]{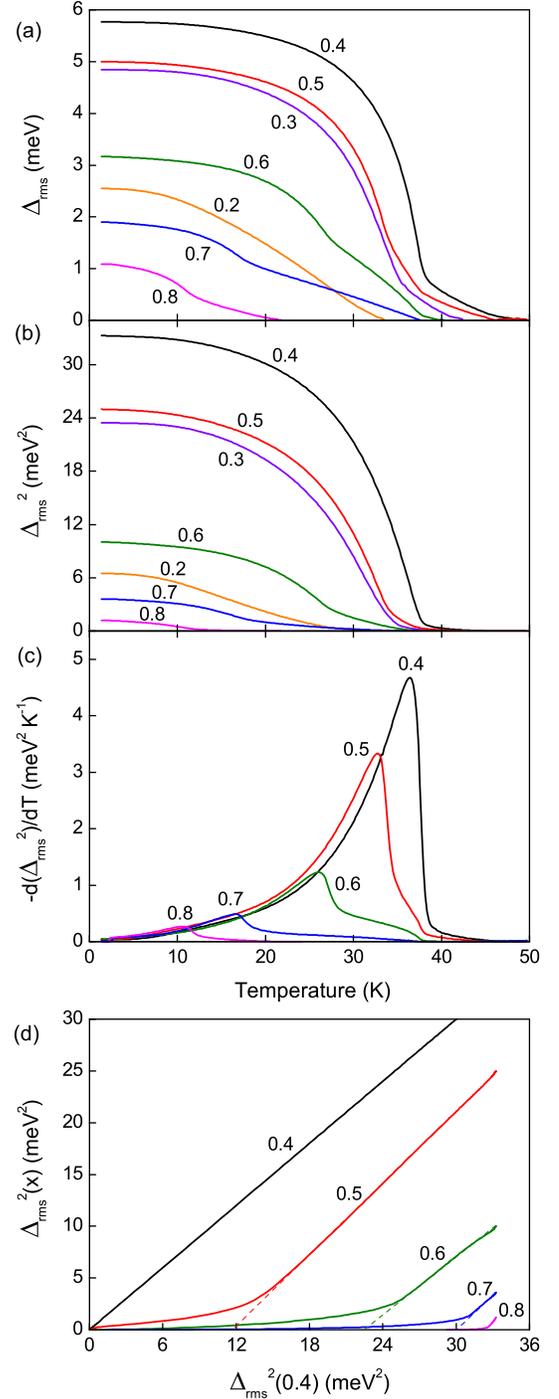}%
\caption{
(Color online) (a) Root-mean-square superconducting gap $\Delta_{rms}$ vs temperature for $x$ = 0.2 to 0.8, calculated from the heat capacity via Eq.~\ref{eq:L6}. (b) $\Delta_{rms}^2$ vs temperature. (c) $-d(\Delta_{rms}^2)/dT$ for $x$ = 0.4 to 0.8 calculated from Eq.~\ref{eq:L7}. The slope of $\Delta_{rms}^2$ below $T_p$ is approximately independent of doping. (d) $\Delta_{rms}^2(x)$ vs $\Delta_{rms}^2(0.4)$ where temperature is an implicit parameter.
} 
\label{DRMSFIG}
\end{figure}

Plots of $\Delta_{rms}(T)$ vs $T$, $\Delta_{rms}^2$ vs $T$, $d(\Delta_{rms}^2)/dT$ vs $T$ obtained via Eq.~\ref{eq:L7}, and $\Delta_{rms}^2(x,T)$ vs $\Delta_{rms}^2(x=0.4,T)$ are shown in Figs.~\ref{DRMSFIG}(a)$-$(d) respectively. The first three plots show a rather abrupt crossover from a more or less conventional $T$-dependence below $T_p$ to an unconventional ``shoulder'' $T$-dependence above $T_p$, the crossover occurring when $\Delta_{rms}(T)$ falls below $\Delta_{rms}(T_p)\sim 1-1.5$meV. The persistence of a finite gap above the shoulder onset $T_{c,s}$, most clearly seen in Fig.~\ref{DRMSFIG}(a), is due to superconducting fluctuations diverging at $T_{c,p}$. 
These plots show several interesting and unusual features. 
Below $T_p$ the slopes $d(\Delta_{rms}^2)/dT$ obtained from Fig.~\ref{DRMSFIG}(c) are almost independent of doping, and give no advance warning of the strongly doping dependent peak heights and shoulders at and above $T_p$. 
Fig.~\ref{DRMSFIG}(d) shows a striking linear relation between $\Delta_{rms}^2(x,T)$ and $\Delta_{rms}^2(0.4,T)$ at all temperatures up to $T_p$, in contrast to the expected negative curvature. This crosses over abruptly at $T_{c,p}$ to a gentle decrease to zero at $T_{c,s}$. In the linear region below $T_p$ 
\begin{equation}
\Delta_{rms}^2(x,T) = a_1(x)\Delta_{rms}^2(0.4,T)-a_2(x)
\label{eq:JL7}
\end{equation}
where $a_1(x)$ = 1.06 and 1.03 for $x$ = 0.5 and 0.6, and the intercept $a_2(x)$ increases approximately linearly with $x$ for $x<0.6$ and more slowly at higher $x$. An unexpected consequence of Eq.~\ref{eq:JL7} is that for $x>0.4$ the main peak transition temperatures $T_{c,p}$ can be predicted simply from a $T$-independent downward shift of $\Delta_{rms}^2(0.4,T)$ by $a_2(x)$. Note that these simple parallel shifts with $x$ are not seen in the curves for $\Delta_{rms}(x,T)$. Interestingly we find that $d(\Delta_{rms}^2)/dT$ is also independent of $H$ below $T_p(H)$ leading to similar parallel downward shifts in $\Delta_{rms}^2$ with $H$ for $x$ = 0.4, 0.5 and 0.6.

We have seen that fluctuations always appear to diverge towards $T_{c,p}$, and conclude that at this temperature the magnitudes of the main peak and knee gaps are close to zero. Assuming that all three gaps are coupled at lower temperatures, coupling to the shoulder gap must therefore weaken as $T$ approaches $T_{c,p}$ leading to a change in the shoulder gap $T$-dependence. However, the fact that the amplitude $A$ of the anomalous $\sqrt{1-T/T_{c,s}}$ dependence in $\gamma(T)$ from the shoulder increases as $x$ goes from 0.7 to 0.5 and $T_{c,s}$ becomes closer to $T_{c,p}$ (Fig.~\ref{GSNFIG}) clearly shows that the main peak and shoulder gaps are not independent above $T_{c,p}$, even though the magnitude of the main peak gap is small. It is possible that residual coupling to fluctuations in the main peak order parameter may be responsible for the anomalous $\gamma(T)=A\sqrt{1-T/T_{c,s}}$ temperature dependence in the shoulder region. The $T$-dependence of $\Delta_{rms}^2$ found above gives $\Delta_{rms}^2(T_p)\sim(2/3N(0))AT_p(1-T_p/T_{c,s})^{1.5}$. If, as observed, coupling to the shoulder gap changes abruptly at a roughly constant value for $\Delta_{rms}^2(T_p)$, this result provides a simple explanation for the divergence of $A$ as $T_{c,s}-T_{c,p}$ decreases.

\section{3-band $\alpha$-model fits}
Guided by the evidence for three distinct gaps in Figs.~\ref{GHVSTFIG} and \ref{DGHTFIG} we have extended the widely-used\cite{KANT,POPOVICH,WEI2,FUKAZAWA} two-band $\alpha$-model\cite{PADAMSEE} to estimate the $T$ = 0 gaps and DOS for the three bands. In the alpha-model the ratio $\alpha=\Delta_0/k_BT_c$ of each gap is an adjustable parameter, and $\Delta(t)=\Delta_0\delta_{BCS}(t)$ where $\delta_{BCS}(t)$ is the normalized BCS gap at $t=T/T_c$\cite{MUHLSCHLEGEL}. For the knee and main peak bands we employ the BCS $T$-dependence $\delta_{BCS}(t)$ , taking the onset temperature of the knee gap to be the same as that of the main peak gap. We integrate both bands over identical Gaussian distributions of onset temperatures, with standard deviation $\sim0.7$K, to simulate the rounding of the main peak. Assuming that a distinct band is responsible for the shoulder, we model its $T$-dependence as follows. 
As discussed above, the $\sqrt{1-T/T_s}$ $T$-dependence of the shoulder implies that near $T_s$, $\Delta_s\propto(T_s-T)^{0.75}$. To incorporate this detail into the $T$-dependence of the shoulder band gap we replace $t=T/T_c$ in the BCS gap function $\delta_{BCS}(t)$ by 
\begin{equation}
y(t)=1-\frac{1-t}{\sqrt{1+t(1-t_0)/(1-t)}}
\label{eq:y}
\end{equation}
$t_0$ is a crossover temperature such that for $t\ll{}t_0$, $\delta_{BCS}(y(t))\rightarrow\delta_{BCS}(t)$ while for $t\gg{}t_0$, $\delta_{BCS}(y(t))\propto{}(1-t)^{0.75}$. 

\begin{figure*}
\centering
\includegraphics[width=\linewidth]{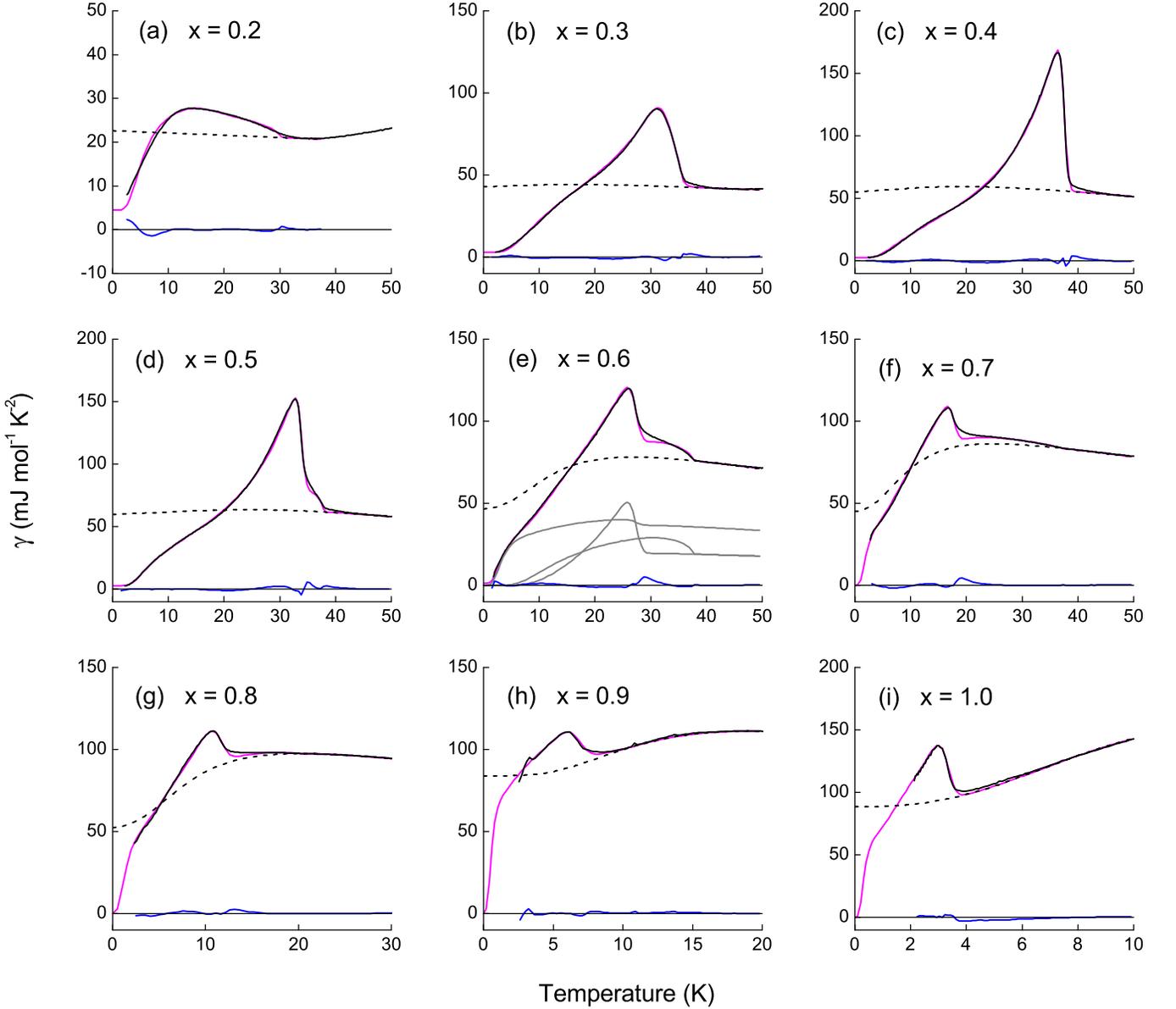}%
\caption{
(Color online) (a) to (h) Normal (dotted line) and superconducting state (magenta line) fits to the zero-field electronic specific heat data (black line) calculated from a 3-band $\alpha$-model, where the shoulder is modeled by a single gap $\Delta_s$ with an initial $(T-T_s)^{0.75}$ dependence. The individual contributions from each band (grey lines) are shown for $x$ = 0.6 in panel (e). Residuals (blue line) are also shown.
}
\label{GFITSFIG}
\end{figure*}

Three-gap fits are shown in Fig.~\ref{GFITSFIG} for $x=0.2-0.8$. For $x\leq0.4$ the fits were made by following the doping dependence of the fit parameters down from higher $x$, with the shape of the knee and main peak below $T_p$ providing good constraints on the range of possible values. For $x$ = 0.9 and 1.0, where no clear evidence for a shoulder is seen and the knee is below our base temperature, a two-gap fit has been applied. Although $\gamma(T)$ for these two samples is still large at 2K, our plots of $S/T$ in Fig.~\ref{ENTROPYFIG} and published data for $\gamma(T)$ below 2K for $x$ = 1\cite{KIM} show that $\gamma_n(0)$ is small in each case. We note that there is considerable evidence for nodes on the Fermi surface of KFe$_2$As$_2$\cite{HASHIMOTO2,REID}, but because the knee gap is small this  effect does not alter our results  significantly.
Overall, the quality of the fits using this model is excellent. The region immediately above the main peak is not quite reproduced since we have not included the fluctuation component diverging towards $T_p$ from above in the fits.

\begin{figure}
\centering
\includegraphics[width=\linewidth]{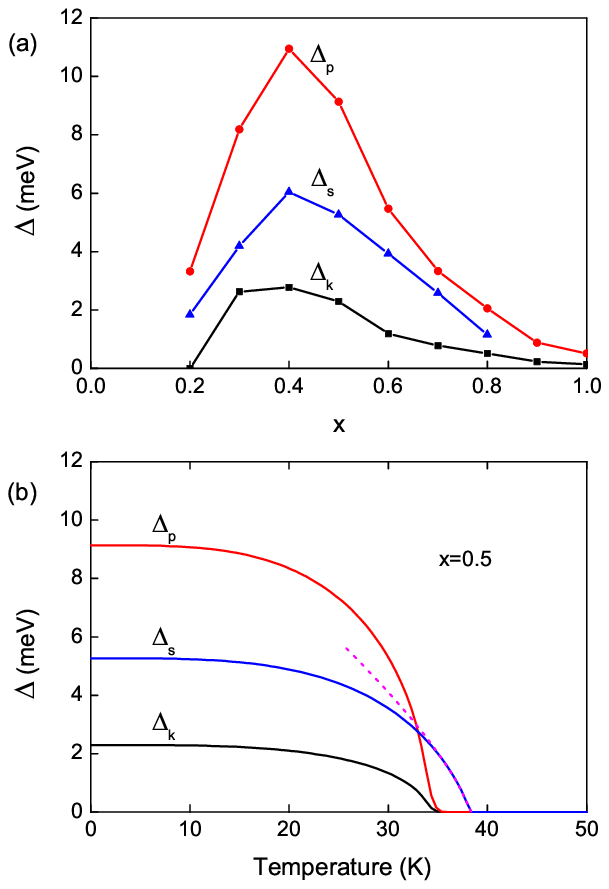}%
\caption{
(Color online) (a) Doping dependence of the knee, main peak and shoulder band superconducting gaps extracted from the fits in Fig.~\ref{GFITSFIG}. (b) Temperature dependence of the knee, main peak and shoulder band gaps employed in the three-band $\alpha$-model fit for $x$ = 0.5. The dotted line shows the initial $(1-t)^{0.75}$ dependence of the shoulder gap.
} 
\label{DVSXFIG}
\end{figure}

Figure~\ref{DVSXFIG}(a) shows the systematic doping dependence of the SC gap magnitudes for each band extracted from the fits. All three gaps show a roughly parabolic doping dependence between $x$ = 0.2 and 0.6 with a maximum near $x$ = 0.4, before tailing off more gradually at higher doping. The $T$-dependence of the three gaps for $x$ = 0.5 is shown in Fig.~\ref{DVSXFIG}(b), which displays behaviour typical of the dopings where a shoulder anomaly is present. The knee and main peak gaps can be understood in terms of the two-coupled-gap model proposed by Suhl \textit{et al}.\cite{SUHL} and Kogan \textit{et al.}\cite{KOGAN}. When the interband coupling in this model is sufficiently large, both gaps approach $T_c$ smoothly, as is the case for $\Delta_k$ and $\Delta_p$ leading to the broad specific heat anomaly at $T_k$ and the sharp peak at $T_p$ discussed above. As discussed in the previous section, the shoulder gap has an unconventional $T$-dependence near $T_{c,s}$. It remains to be seen if a three-band coupled-gap model can give rise to such an effect. Sign changes in the gaps and frustration effects may play a key role, and we would welcome input from theorists on this matter. Dias and Marques\cite{DIAS} have already demonstrated some of the unusual $T$-dependences that can arise from a frustrated multiband model. A further curious result of our analysis is that the shoulder gap $\Delta_s(0)$ appears to be smaller than the main peak gap $\Delta_p(0)$, even though $T_{c,s} > T_{c,p}$ (see Fig 21(b)). Interestingly it has been demonstrated that such behaviour can arise from the self-consistent BCS gap equation for $d_{x^2-y^2}+id_{xy}$\cite{GHOSH98} or $d_{x^2-y^2}+is$\cite{GHOSH99} symmetries. Moreover an intermediate $s+is$ symmetry has been proposed in Ba$_{1-x}$K$_x$Fe$_2$As$_2$ for $x$ between 0.4 and 1.0\cite{MAITI,STANEV,TANAKAY2}. We therefore propose that this feature may signify the presence of mixed or unusual competing order parameters.

\begin{figure}
\centering
\includegraphics[width=\linewidth]{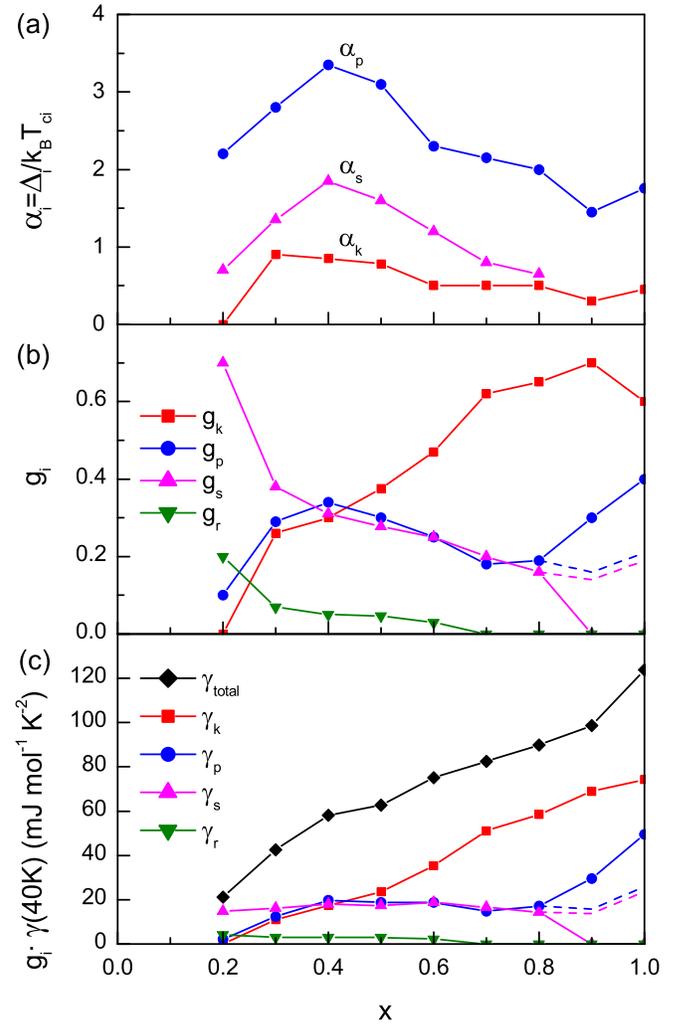}%
\caption{
(Color online) The parameters extracted from the fits in Fig.~\ref{GFITSFIG}. (a) Ratio of superconducting gap magnitude to onset temperature for the knee (\textit{k}), main peak (\textit{p}) and shoulder (\textit{s}) bands. (b) The fraction $g_i$ of $\gamma_n(T)$ contributed by each band. $g_r$ is an ungapped fraction of $\gamma_n$, included to model the residual $\gamma(0)$. (c) The band fractions multiplied by $\gamma(40K)$. Dotted lines in (b) and (c) show the behaviour if the main peak and shoulder bands contribute roughly equally to $\gamma_n$.
} 
\label{ALPHAFIG}
\end{figure}

In addition to the SC gap magnitudes, the three-band fits reveal the doping dependence of the fractions of the DOS contributed by each band $g_k$, $g_p$ and $g_s$ and their relative contributions  $g_i\gamma_n(40\mathrm{K})$ to the normal state electronic term $\gamma_n(40\mathrm{K})$. These quantities are shown in Fig.~\ref{ALPHAFIG}. Above $x$ = 0.2, the ``knee'' contribution $g_k\gamma(40\mathrm{K})$ increases steadily, characteristic of a hole-like band and this band contributes more than 60$\%$ of the DOS at high doping. Although we are unable to resolve a third gap for $x$ = 0.9 and 1 it is possible that there are three gaps across the entire doping range. If this is the case and if the shoulder and main peak bands contribute roughly equally for $x$ = 0.9 and 1, we obtain the dashed lines shown in Fig.~\ref{ALPHAFIG}. 
Values of $g_i\gamma_n(40\mathrm{K})$ for the main peak and shoulder bands are almost identical and relatively doping independent up to at least $x$ = 0.8, though their contribution to the total DOS decreases with $x$. 

It is important to assess the reliability of the parameters deduced from the fits shown in Figs.~\ref{DVSXFIG} and \ref{ALPHAFIG}. For the ``knee'' band, the value of the gap $\Delta_k(0)$ and the normal state DOS fraction $g_k$ can be determined with confidence from the temperature and magnitude of the knee in $\gamma(T)$. The values of $\Delta_p(0)$ and $\Delta_s(0)$ for the main peak and shoulder anomalies depend on the assumption that distinct bands are responsible for the peak and shoulder features in $\gamma(T)$. On that assumption, fitting the strong positive curvature in $\gamma(T)$ below the peak with the $\alpha$-model leads to the large (strong coupling) values for the gap $\Delta_p(0)$ and for $\alpha_p$ shown in Figs.~\ref{DVSXFIG} and \ref{ALPHAFIG}(a). It should be noted however that $\alpha$-model fits focusing primarily on the peak region may overestimate $\Delta(0)$ and $\alpha$ for strong coupling\cite{PADAMSEE}. This overestimate would have little effect on the fit to $\gamma(T)$ at lower temperatures (Fig.~\ref{GFITSFIG}) since $\gamma(T)$ is small and insensitive to $\Delta(T)$ when $k_BT\ll\Delta(T)$. The reliability of the shoulder gap $\Delta_s(0)$ depends somewhat on the validity of the interpolation from the BCS to shoulder $T$-dependences for $\Delta_s(T)$ discussed above, and is therefore difficult to assess. We are confident however that the values of the DOS fractions $g_i$ and the contributions to the normal state $\gamma_n(40\mathrm{K})$ for all three bands are reliable.

\section{Comparison with Band Structure.}
\label{secBANDSTRUC}
Studies of the Fermi surface (FS) of Ba$_{1-x}$K$_x$Fe$_2$As$_2$ by Shubnikov-de Haas oscillations and angle-resolved photoemission spectroscopy have yielded the following observations. In the SDW phase the Fermi surface consists of small pockets of hole- and electron-like character\cite{ANALYTIS,TERASHIMA2}. These are believed to arise from band folding combined with finite $k_z$ corrugation. In the tetragonal phase the FS is composed of three concentric hole sheets at the $\Gamma$ point\cite{ZHANG} (also seen in electron-doped Ba(Fe$_{1-x}$Co$_x$)$_2$As$_2$\cite{BROUET}), and a propeller-like structure at the $M$ point made up of hole-like blades surrounding an electron-like center\cite{ZABOLOTNYY,ZABOLOTNYY2,SATO2}. With increasing $x$ the hole-like surfaces expand, and the electron-like surface shrinks before disappearing near $x$ = 1.0\cite{SATO2}. A small superconducting gap is observed on the outer $\Gamma$ pocket, while larger gaps are detected on the inner $\Gamma$ pocket(s) and $M$ pockets\cite{DING3,EVTUSHINSKY}. With these observations in mind it is possible to assign our bands to particular FS sheets. The knee band has a small gap and increases with doping, behaviour consistent with the outer $\Gamma$ hole pocket. Note that this quasi-2D sheet has a large value of $m^*$ \cite{TERASHIMA}. The shoulder band has a larger gap, and if this band vanishes above $x$ = 0.8 it would be consistent with the electron pocket at the $M$ point. The main peak band has the largest gap consistent with inner $\Gamma$ hole pockets. The increase in $g_p$ above $x$ = 0.8, shown in Fig.~\ref{ALPHAFIG}, suggests that this band also includes contributions from the hole-like blades at the $M$ point. However, it is also possible that for $x$ = 0.9 and 1.0 both the main peak and shoulder bands continue to make an approximately equal contribution to $\gamma_n$ (dashed lines in Fig.~\ref{ALPHAFIG}). At $x$ = 0.4, $T_c$, $\Delta_{rms}$ and the superconducting condensation energy are maximal (Figs.~\ref{U0FIG} and \ref{DRMSFIG}(a)). At this particular doping we note that the three bands contribute equal fractions to the DOS. If $\bigtriangledown \epsilon_k$ is similar for each band, this would imply that the hole and electron pockets are roughly the same size, and support the hypothesis that $T_c$ in this system is governed by the degree of Fermi surface nesting\cite{MAZIN,KUROKI,KUROKI2}. 

\begin{figure}
\centering
\includegraphics[width=\linewidth]{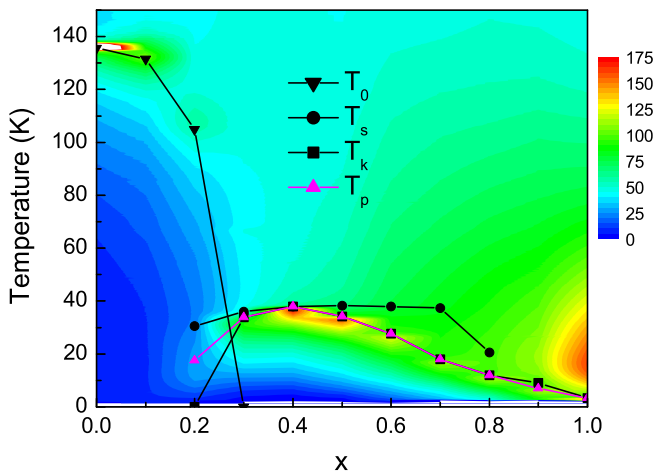}%
\caption{
(Color online) The doping dependence of the magneto-structural transition temperature $T_0$, and onset temperatures of the SC gaps of each band, overlaid on a false-color plot of the electronic specific heat. Subscripts \textit{k}, \textit{p} and \textit{s} correspond to the knee, main peak and shoulder features respectively.
} 
\label{FCFIG}
\end{figure}

In Fig.~\ref{FCFIG} we show a temperature-doping phase diagram comprised of the magneto-structural transition temperature $T_0$ and onset temperatures of the three SC gaps, overlaid on a false color plot of $\gamma(x,T)$.
For $x<0.3$, the superconducting phase competes with, and ultimately succumbs to, a decrease in spectral weight due to increasing gapping in the SDW phase. An analogous situation occurs in the pseudogap phase of the high-$T_c$ cuprates\cite{STOREYPG}. At higher dopings, $x>0.4$, the superconducting transition temperatures and condensation energies fall in the presence of an increasing normal-state $\gamma$. Interestingly, in overdoped cuprates where $T_c$ is less than optimal, $\gamma$ is also large\cite{TALLON6}. The observed fall in $T_c$ despite the presence of a growing DOS indicates that the DOS might not be the dominant factor governing $T_c$. It is possible in the case of BKFA that the fall in $T_c$ with $x$ is driven by increasingly poor FS nesting as the hole pockets expand and the electron pocket shrinks.

\section{Summary.}
In summary, using a high-resolution differential technique we have determined the electronic specific heat $\gamma$ of Ba$_{1-x}$K$_x$Fe$_2$As$_2$ with $x$ = 0 to 1.0, from 2K to 380K and in magnetic fields 0 to 13T. In the SDW phase the low-temperature normal-state values of $\gamma_n$ are reduced relative to their values at high temperature by factors of 10, 5 and 2 for $x$ = 0, 0.1 and 0.2 respectively, reflecting partial gapping or reconstruction of the Fermi surface. Near optimal doping $\gamma_n$ is practically $T$-independent. As $x$ increases to 1.0 an increase in $\gamma_n$ with $x$ at low-$T$ is accompanied by a corresponding decrease at high-$T$, consistent with a substantial renormalisation of the effective mass as seen experimentally by de Haas van Alphen studies for $x$ =1 \cite{TERASHIMA}.

In the superconducting state we have observed a new feature. In addition to the well-known knee and peak features that are typically associated with two distinct bands and SC gaps, we have identified a shoulder feature above the main peak with an abrupt onset temperature $T_{c,s}$. Our attempts to explain this feature in terms of doping inhomogeneity fail to withstand rigorous analysis on several levels. In particular, the extent of inhomogeneity implied by the breadth of the shoulder is inconsistent with X-ray diffraction spectra and is contradicted by the consistently sharp transitions of the main peak. Hence we conclude that the samples are spatially homogeneous and instead attribute the shoulder to a third band and SC gap. The anomalous $\sqrt{1-T/T_{c,s}}$ dependence of $\gamma(T)$ in the shoulder region and the absence of a mean field jump at $T_{c,s}$ await theoretical treatment. An analysis of Gaussian fluctuations above the main peak yield superconducting coherence lengths $\xi_{ab}\sim$ 20\AA\ and $\xi_c\sim$ 3\AA. It is possible that the separate onset temperatures of the shoulder and main-peak gaps signify the presence of mixed or unusual competing order parameters.

The doping dependence of the three gaps and bands was extracted via the application of a three-band $\alpha$-model. The SC gaps, $T_c$ and the condensation energy are all maximal at $x$ = 0.4 and the evolution of the bands is consistent with changes in the Fermi surface  observed by ARPES. At this doping the three bands contribute equal fractions to the density of states. 
Finally, the sub-linear magnetic field dependence of $\gamma$ at low $T$ for $x$ = 0.3 to 0.5 points towards an unusual pairing symmetry such as $s\pm$-wave. This important point needs more detailed analysis, though ideally the field dependences should be determined at lower temperatures.

\begin{acknowledgements}
We gratefully acknowledge funding from the Engineering and Physical Sciences Research Council U.K. (grant number EP/G001375/1) and the Swiss National Science Foundation pool MaNEP.
\end{acknowledgements}

\appendix*
\section{Estimate of the doping inhomogeneity required to explain the shoulder feature.}
To explain the shoulder feature on this interpretation requires a broad spread of local doping with probability $P(x)$ between $x_1$ and $x_2$. The corresponding spread of local $T_c$'s has a probability distribution $P(T_c)= P(x)/|dT_c/dx|$ between $T_{c,min}$ and $T_{c,max}$. For a parabolic dependence $T_c(x)=T_m[1-((x-x_m)/w)^2]$ peaking at $x_m$, we have $|dT_c/dx|=(2T_m/w^2)|x-x_m|$ and $P(T_c)=P(x)/\sqrt{1-T_c/T_m}$. 
The contribution to the specific heat coefficient $\gamma_{sn} = \gamma_s -\gamma_n$ is given by
\begin{equation}
\gamma_{sn} = \int{\gamma_{sn}(T/T_c)P(T_c)dT_c}	
\label{eq:A1}
\end{equation}
where for simplicity we assume that the function $\gamma_{sn}(T/T_c)$ representing the unbroadened mean-field transition has a discontinuous jump $\Delta\gamma$ at $T_c$ and is zero for $T>T_c$. Since only those regions with local $T_c$ greater than $T$ contribute to the integral in Eq.~\ref{eq:A1}, the lower limit is $T_{c,min}$ if $T<T_{c,min}$ and $T$ if $T_c>T_{c,min}$. If $x_m$ for optimum $T_c$ lies within the range $x_1$ to $x_2$ and $T_c(x)$ is parabolic through $x_m$, then $P(T_c)= P(x)/\sqrt{1-T_c/T_m}$, which diverges at $T_m$. Close to $T_m$ Eq.~\ref{eq:A1} gives $\gamma_{sn}(T)= \Delta\gamma2wP(x_m)\sqrt{1-T/T_m}$ in agreement with the observed $T$-dependence for $x$ = 0.5 and 0.6. If $x_m$ lies outside the range $x_1$ to $x_2$ and if $P(T_c)\propto(1-T_c/T_{c,max})^n$, Eq.~\ref{eq:A1} gives $\gamma_{sn}(T)\propto(1-T/T_{c,max})^{n+1}$ where we identify $T_{c,max}$ with $T_{c,s}$.	In all cases we expect a significant reduction in slope $d\gamma_{sn}/dT$ when $T$ increases through $T_{c,min}$ due to the reducing superconducting volume fraction. This is not observed in our data.

The range of the broad distribution can be estimated as follows. Ignoring the small variation through the ``knee'' feature, the slope $d\gamma_{ns}/dT$ increases continuously up to the main peak (Fig.~\ref{DGDTFIG} \& Fig.~\ref{GSNVSTFIG}). Since there is no reduction of slope anywhere below $T_p$ we conclude that $T_{c,min}\sim T_p$ and thus $x_1\sim x_0$, the nominal doping. For samples exhibiting an $A\sqrt{1-T/T_m}$ dependence between $T_p$ and $T_m$ the range must extend at least from the nominal $x_0$ to $x_m$ with a significant probability $P(x_m)\propto A$. Unless $P(x)$ drops discontinuously to zero at $x_m$, which seems unlikely, we require that the spread extends an equal range on the opposite side of $x_m$ to avoid a slope change above $T_p$. So $T_c(x_1)\sim T_c(x_2)$, making a minimum total range $\Delta x\sim2|x_0-x_m|$. 

Thus to explain our data the doping inhomogeneity would have to extend over the range $x_0$ to $2x_m-x_0$ with width $\Delta x\sim2|x_0-x_m|$. Not only is this range very wide, but it only appears to exist on one side of the nominal doping $x_0$ (always towards $x_m$). Contrary to the expectation that the distribution $P(x)$ should be reasonably symmetric around $x_0$ and have a mean value $x=x_0$, we find instead that it would have to be very asymmetric, extending from $x_0$ to well beyond $x_m$, with a mean doping substantially different from $x_0$.

\end{document}